\documentclass[journal,onecolumn]{IEEEtran}
\usepackage{amsmath,amsfonts,amsthm,amssymb}
\usepackage{mathrsfs}
\usepackage{algorithmic}
\usepackage{algorithm}

\usepackage[caption=false,font=normalsize,labelfont=sf,textfont=sf]{subfig}
\usepackage{tikz}
\usetikzlibrary{matrix,positioning,decorations.markings,shapes.geometric}
\usepackage{pgfplots}
\pgfplotsset{compat=1.18}
\usepackage{color}
\usepackage{graphicx}
\usepackage{cite}
\usepackage{booktabs}  % package for \toprule and bottomrule

\theoremstyle{plain}                         
\newtheorem{theorem}{Theorem}
\newtheorem{lemma}[theorem]{Lemma}
\newtheorem{coro}[theorem]{Corollary}

\theoremstyle{definition}                   
\newtheorem*{definition}{Definition}
\newtheorem*{construction}{Construction}

\theoremstyle{remark}
\newtheorem*{remark}{Remark}
\newtheorem{example}{Example}

\newcommand{\be}{\mathbf{e}}
\newcommand{\bu}{\mathbf{u}}

\newcommand{\bL}{\mathbf{L}}

\newcommand{\sJ}{\mathcal{J}}

\newcommand{\sA}{\mathcal{A}}
\newcommand{\sV}{\mathcal{V}}
\newcommand{\sE}{\mathcal{E}}
\newcommand{\sB}{\mathcal{B}}
\newcommand{\FF}[1]{\mathbb{F}_{\!#1}}
\newcommand{\sF}[1]{\mathbf{F}_{#1}}
\newcommand{\sP}[1]{\mathbf{P}_{#1}}
\newcommand{\sK}{\mathbf{K}}
\newcommand{\mV}[1]{\mathcal{R}_{#1}}

\begin{document}

\title{Regenerating codes with minimal disk I/O cost achieving optimal tradeoff between storage and repair bandwidth}

\author{Minhan Gao and Kenneth W. Shum \\
School of Science and Engineering\\
The Chinese University of Hong Kong, Shenzhen \\
Email: minhangao@link.cuhk.edu.cn, wkshum@cuhk.edu.cn
}

\maketitle

\begin{abstract}
Regenerating codes achieve the fundamental tradeoff between storage efficiency and repair bandwidth in distributed storage systems. Beyond these two parameters, disk I/O cost is an important measure of repair efficiency, capturing the number of stored packets accessed at the helper nodes during repair. A repair scheme is access-optimal if each helper reads exactly as many packets as it transmits, and is help-by-transfer if each helper sends stored packets directly to the newcomer without local computation. In this paper, we study functional repair of a single node failure in the regime where all surviving nodes participate as helpers. We introduce a framework based on signal flow graphs and gammoids, which separates the combinatorial structure of the repair process from its linear-algebraic realization. Within this framework, we construct functional-repair regenerating codes that attain every point on the optimal storage-bandwidth tradeoff curve. The proposed codes are help-by-transfer and access-optimal. Moreover, they operate over a fixed finite field and preserve the data-recovery property under an arbitrarily long sequence of repairs.
\end{abstract}


\section{Introduction}
The rapid growth of large-scale data-intensive applications has made reliability and efficiency central design objectives in distributed storage systems. Modern storage systems, such as the Google File System~\cite{GFS} and Windows Azure Storage~\cite{Azure,CHuang12}, distribute data across multiple storage nodes and must tolerate frequent node failures. A standard approach to fault tolerance is erasure coding. In particular, maximum distance separable (MDS) codes provide optimal storage efficiency: an encoded file stored across $n$ nodes can be recovered from any $k$ of them. Reed--Solomon codes and their variants are among the most widely used MDS codes in practice. However, although MDS codes are optimal from the viewpoint of storage overhead, repairing a failed node may require downloading an amount of data much larger than the amount stored in the failed node. This repair bottleneck has motivated extensive work on low-bandwidth repair of MDS and Reed--Solomon codes; see, for example, \cite{guruswami2016repairing,dau2018repairing,tamo2017optimal,DuursmaDau,Dinh22,liu2024formula}. Regenerating codes take a complementary viewpoint by treating repair bandwidth as a fundamental design parameter.

%A key advantage of MDS codes lies in their high storage efficiency and optimal error-correcting capability. However, MDS codes are not very amenable to node repair, incurring substantial network resources to regenerate a storage node. In other words, recovering data from failed nodes requires accessing and transferring significantly more data across the network than the amount of lost data. Several research efforts \cite{guruswami2016repairing,dau2018repairing,tamo2017optimal, Dinh22, liu2024formula} have been dedicated to reducing the repair bandwidth of RS codes. Another approach is to consider a formulation of RS code using a Cauchy matrix. Utilizing the special structure of Cauchy matrices, we can further reduce the repair bandwidth~\cite{DuursmaDau, Longhair}.

Regenerating codes were introduced by Dimakis \emph{et al.}~\cite{DGWR2010} to characterize the tradeoff between storage efficiency and repair bandwidth in distributed storage systems. In the standard $(n,k,d)$ model, a file of size $B$ is encoded and stored across $n$ nodes, each storing $\alpha$ symbols. The file must be recoverable from any $k$ nodes. When a node fails, a newcomer contacts any $d$ surviving helper nodes and downloads $\beta$ symbols from each helper, for a total repair bandwidth $\gamma=d\beta$. By modeling the evolution of the storage system using an information flow graph and applying the max-flow/min-cut bound from network coding theory, \cite{DGWR2010} derived the fundamental cut-set bound
\begin{equation}
 B \leq \sum_{i=1}^k \min\{\alpha, (d-i+1)\beta \}.
 \label{eq:tradeoff}
\end{equation}
Furthermore, this inequality is tight, with equality  achieved through network coding~\cite{Wu09}. 
The points that achieve equality in this bound delineate the optimal tradeoff between the storage per node $\alpha$ and the repair bandwidth $\gamma$. The two extreme points of this tradeoff are the minimum-storage regenerating (MSR) point and the minimum-bandwidth regenerating (MBR) point.

The original information-flow-graph model allows {\em functional repair}. Under functional repair, the new node need not reproduce exactly the content of the failed node; it is only required that, after repair, the data reconstruction property remains valid. Thus, after any sequence of repairs, the file must still be recoverable from any $k$ active nodes. This flexibility is closely related to network coding, where intermediate nodes may perform computations on the data they receive. It is proved in~\cite{Wu10} that capacity-achieving functional-repair regenerating codes exist under general conditions, and network-coding methods can achieve the cut-set bound over sufficiently large finite fields. 

%Regenerating codes  were introduced in~\cite{DGWR2010} to explore the  tradeoff between storage efficiency and repair bandwidth. The authors of~\cite{DGWR2010} model a distributed storage system by an {\em information flow graph.} By analyzing the network capacity using the max-flow-min-cut bound from network coding theory, a fundamental bound on repair bandwidth is derived.
%A key feature of network coding is that computations and calculations are enabled in intermediate relay nodes in a network. Hence, in this model, a node that helps recovering a failed node may access all stored data, compute and send some data to the new node just for the purpose of regeneration. This mode of repair is called {\em functional-repair}, since the content of the new node after the repair may not be the same as the content before the failure, but the functionality of the distributed storage system is maintained.
%Substantial research has focused on this paradigm (e.g. \cite{cadambe2010distributed,  rashmi2011optimal, suh2011exact}).

%Another repair mode for regenerating codes is exact repair (ER) \cite{wu2009reducing,shah2010explicit}, where the newcomer reconstructs data identical to the failed node. However, it is proved in \cite{Tian2014} that in general exact-repair regenerating codes cannot attain the operation points on the fundamental tradeoff curve between repair bandwidth and storage per node.

A more restrictive but practically attractive repair model is {\em exact repair}, in which the newcomer reconstructs precisely the data that was stored in the failed node. Exact repair keeps the code structure invariant over time and has been extensively studied, especially at the MSR and MBR points; see, for example, \cite{wu2009reducing,shah2010explicit,SK11,rashmi2011optimal,suh2011exact,SRKR10f}. However, exact-repair regenerating codes do not in general attain the functional-repair cut-set bound at all interior points of the storage-bandwidth tradeoff; a strict separation for the $(4,3,3)$ case was established in~\cite{Tian2014}. This motivates the study of explicit constructions of functional-repair regenerating codes, especially in the interior points of the tradeoff curve~\cite{hollmann2025family}.

In addition to network repair bandwidth, disk I/O is another important performance metric. In large-scale storage systems, each node may store a large amount of data, and the time required merely to read data from disk can dominate the repair latency. Disk reads can also interfere with normal service at the helper nodes~\cite{Khan11,Khan12,WTB11,TWB12,Shah13}. This has led to the study of repair schemes with low access complexity. In the \emph{help-by-transfer} (HBT) model, each helper reads a subset of its stored symbols and transfers them directly to the newcomer without performing local computation; the newcomer may then compute its new stored symbols from the received data. A stronger form is \emph{repair-by-transfer} (RBT), or uncoded repair, in which the repair process avoids coding operations in the transfer stage~\cite{ElRouayheb10, SRKR12}. A repair scheme is called access-optimal if the number of symbols read from each helper equals the number of symbols transmitted by that helper. Thus, in an access-optimal HBT scheme, the disk-read cost matches the repair bandwidth.

%Beside the repair bandwidth, disk I/O cost is another performance metric. Due to the large volume of the data stored in a node, merely reading the data already incurs delay. Also, the normal operation of a storage node is interrupted by the disk reading operations. Another limitation is the computational power of storage devices, which may not be able to perform complicated calculations efficiently. The disk I/O cost is another performance metric in addition to repair bandwidth. To address this issues, two repair models were proposed. The first one is called help-by-transfer (HBT) \cite{ramkumar2022codes} model, each surviving node reads only a subset of its stored data and transfers the data directly to the newcomer without computation. The new node will be compute the new content by combining the received data appropriately. The other method of repair with is called ``repair by transfer'' (RBT), in which both the helper nodes and the new node requires no computation~\cite{SRKR12}, which is also termed uncoded repair \cite{ElRouayheb10}. In both methods, each helper node reads exactly as many data packets as it transmits, with the code reaching the optimal tradeoff.

While the MSR and MBR points have received substantial attention, the interior points of the storage-bandwidth tradeoff are less well understood. Several constructions for interior points have been developed, many of which are based on exact repair or related algebraic and graph-theoretic techniques~\cite{senthoor2015improved,elyasi2016determinant,elyasi2020cascade,duursma2021multilinear,patra2022interior}. However, as noted above, exact repair cannot in general achieve the full functional-repair tradeoff. On the functional-repair side, existence results and randomized network-coding constructions are known~\cite{Wu10}, and practical or structured functional-repair schemes have been investigated in \cite{NCCloud,HLS13,SH12,liu2015general}. Nevertheless, explicit functional-repair constructions that simultaneously achieve the optimal tradeoff, require only help-by-transfer, are access-optimal, and support an unbounded number of repairs remain limited. Recently, Hollmann~\cite{hollmann2025family} presented a family of optimal linear functional-repair codes realizable as HBT codes, but the construction is restricted to the case $n=k+1$.

%The two extremes of the tradeoff between storage and repair bandwidth are minimum-storage regenerating (MSR) codes and minimum-bandwidth regenerating (MBR) codes.
%While the majority of existing research focuses on MSR and MBR points, interior points receive substantially less attention \cite{senthoor2015improved,elyasi2016determinant,elyasi2020cascade,duursma2021multilinear,patra2022interior}. Results on interior points predominantly assume exact repair, despite its proven inability to achieve the optimal efficiency-bandwidth tradeoff \cite{SRKR12}. For functional-repair approaches, we refer to \cite{wu2010existence} which proves the existence of functional-repair regenerating codes achieving the optimal tradeoff for sufficiently large fields. Functional MSR codes were initially proposed in \cite{hu2012nccloud} and later analyzed in \cite{HLS13}. The heuristic algorithm of~\cite{liu2015general} targets functional repair on the full tradeoff curve but does not guarantee optimal repair bandwidth. Recently, \cite{hollmann2025family} presented a family of linear optimal functional-repair codes realizable as HBT, with the construction restricted to $n=k+1$.

In this paper, we introduce a new framework for functional-repair regenerating codes based on signal flow graphs and gammoids. A gammoid is a matroid whose independent sets are characterized by vertex-disjoint linkings in a directed graph~\cite{Mason72,Perfect69}. This viewpoint allows us to separate the repair problem into a combinatorial part and a linear-representation part. The combinatorial part determines how a new stage is appended to the signal flow graph so that the required linking conditions continue to hold after repair. The algebraic part then chooses coding coefficients so that the corresponding gammoid independences are represented as linear independences over a finite field. In this way, the recovery condition from any $k$ active storage nodes is maintained throughout the repair process.

%Most prior constructions are described directly using matrices and vectors. In this work we adopt a different approach, reformulating the problem using signal flow graphs and gammoids. Each repair requires the newcomer to reconstruct its content from the packets transmitted by the helpers, so that the original data can be recovered from any $k$ nodes. By introducing the gammoids on the signal flow graph, the repair process separates into the combinatorial part and the linear representation part. That is, each helper determines the packets transmitted to the newcomer according to the combinatorial independence introduced by the gammoids, which is realized by appending a new stage to the signal flow graph that satisfies the linking conditions. The newcomer then represents this combinatorial independence as linear independence by choosing appropriate coefficients, preserving the recovery property.

The main contributions of this paper are as follows.

\begin{itemize}
    \item We formulate functional repair for distributed storage systems through signal flow graphs and gammoids. This gives a combinatorial criterion for preserving the data-reconstruction property after each repair and provides a systematic way to design functional-repair regenerating codes.

    \item For the regime $d=n-1$, the proposed functional-repair  procedure is help-by-transfer and access-optimal, attaining every point on the optimal storage-bandwidth tradeoff curve. Each helper simply transfers selected stored symbols to the newcomer without local computation, and the number of symbols read from each helper equals the number of symbols transmitted.
    Under these stringent requirements, our construction can achieve the cut-set bound for functional-repair regenerating codes. 

    \item Our proposed construction is a linear code over a fixed finite field and supports an unbounded sequence of repairs through causal, finite-memory operations. The field size and the per-repair cost are independent of the number of previous repairs. Although the information flow graph may grow unboundedly over time, each repair operation depends only on a finite memory window. Consequently, the complexity of our algorithms remains bounded as the system evolves.
    
\end{itemize}

%Our contributions consist of two parts. The first is the framework described above. To the best of our knowledge, this perspective is new for regenerating codes. Second, in the regime $d=n-1$, within this framework we construct a functional-repair code that attains all points on the optimal tradeoff curve, and the resulting code is both HBT and access-optimal. To the best of our knowledge, no prior construction attains this combination at once. Moreover, our construction operates over a finite field of fixed size, and tolerates an unbounded number of repairs. The cost of each repair does not grow with the number of preceding repairs.

Table~\ref{tab:comparison} summarizes the comparison between this paper and related work. Wu~\cite{Wu10} used the path-weaving technique to establish the existence of functional-repair regenerating codes attaining the fundamental tradeoff between storage per node and repair bandwidth. The disk I/O cost, however, was not considered in~\cite{Wu10}.  Regenerating codes with minimum disk I/O cost at the MSR point were studied in~\cite{HLS13,SH12}. Hollmann~\cite{hollmann2025family} constructed access-optimal functional-repair regenerating codes for the special case $k=d=n-1$. In contrast, this paper considers arbitrary $k$ with $d=n-1$ and achieves the entire functional-repair tradeoff curve with help-by-transfer and access-optimal repair. To the best of our knowledge, this is the first construction that, for $d=n-1$, simultaneously achieves all points on the functional-repair storage-bandwidth tradeoff curve, is help-by-transfer and access-optimal, and supports an unbounded number of repairs over a fixed finite field.

\begin{table}[ht]
	\centering
	\caption{Comparison with Prior Functional-repair Regenerating Codes}
	\label{tab:comparison}
	\begin{tabular}{c c c c c}
		\toprule
        \textbf{Regenerating codes} & \textbf{Parameters} & \textbf{Tradeoff points}&\textbf{HBT and access-optimal}& \textbf{Field size}\\ \midrule
        Wu \cite{Wu10} & $(n,k,d)$ & All& No & $\geq \binom{n\alpha}{B}B$\\
        Hu {\em et al.} \cite{HLS13} \cite{SH12} & $(n,k,d)$ & MSR point & Yes & Large enough\\
        Hollmann  \cite{hollmann2025family} & $(n,k=n-1,d=n-1)$& All& Yes & $\geq k-1$\\
        This work & $(n,k,d=n-1)$ & All& Yes & $> \binom{n\alpha}{B}-\binom{(n-1)\alpha}{B}$\\
        \bottomrule
	\end{tabular}
\end{table}

The remainder of this paper is organized as follows. Section~\ref{sec: problem} formulates the repair problem and defines the performance metrics, and Section~\ref{sec:linear_regenerating_code} models the HBT functional-repair process as a discrete-time linear dynamical system. Section~\ref{sec: matroid} reviews the matroid and gammoid background. Section~\ref{sec: graph} introduces the signal flow graph and the gammoids defined on it. 
Section~\ref{sec: reduction} sets out the construction strategy, decomposing the code construction into a combinatorial layer and a linear representation layer. Once the linear representability of the gammoids is established, the code construction is reduced to the construction of a signal flow graph satisfying the required conditions. Section~\ref{sec:packet_selection} provides the explicit construction of the signal flow graph. Section~\ref{sec: linear} provides the proof of the main theorem in this paper. Some concluding remarks are given in Section~\ref{sec:conclusion}.

\section{Problem Formulation for Regenerating Codes with Functional Repair and Help by Transfer}
\label{sec: problem}

Consider the problem of encoding a data file in a distributed storage system consisting of $n$ storage nodes. We use a {\em packet}, also referred to as a {\em symbol}, as the basic unit for measuring the amount of data. Our goal is to encode and distribute a data file containing $B$ packets across the $n$ storage nodes, such that each storage node stores $\alpha$ packets. The encoding is designed so that a data collector can recover the original data file by downloading data from any $k$ nodes at any time. This property is referred to as the $(n,k)$ {\em recovery property}, and the parameter $k$ is called the {\em recovery degree}. For any positive integer $a$, we denote by $[a]$ the set $\{1,2,\dots,a\}$. The notation used in this paper is summarized in Table~\ref{tab:notations_summary}.

\begin{table}[ht]
	\centering
	\caption{Summary of Notation}
	\label{tab:notations_summary}
	\begin{tabular}{cl}
		\toprule
		\textbf{Notation} & \textbf{Description} \\ \midrule
		$n$    & Number of storage nodes \\
		$k$    & Recovery degree \\
		$d$    & Repair degree (in this paper $d=n-1$) \\
		$B$    & File size \\
		$\alpha$ & Storage capacity per node \\
		$\beta$ & No. of packets sent from a helper to the newcomer \\
		$\ell$  & Index of the operating point on the tradeoff curve\\
		$t$    & Stage index \\
		$q$    & Finite field size \\
		$\FF{q}$ & Finite field of size $q$ \\
		$\mathbf{m}$  & A vector over $\FF{q}$ representing the data file \\
		$Q_b$ & The $b$-th source packet, $b\in[B]$ \\
		$\mathbf{u}_b$ & The $b$-th standard basis vector of $\FF{q}^B$ \\
		$S_t(i,j)$ & The $j$-th packet in the $i$-th node at stage $t$, \\
		& represented as an element in $\FF{q}$ \\
		$\mathbf{e}_t(i,j)$ & Global encoding vector of $S_t(i,j)$ \\
		$b_t(i,j)$ & Local encoding coefficients at stage $t$ \\
		$v_t(i,j)$ & Vertex corresponding to $\be_t(i,j)$ in the signal flow graph\\
		$F_t$ & Index of the failed node at stage $t$ \\
		$p_t(i)$ & The choice function of node $i$ at stage $t$ \\
		$\mathbf{K}$ & Set of $k$ nodes $\{i_1,\dots,i_k\}$ in the $(n,k)$ recovery property \\
		$\sJ$ & Index set $\{(i,j)\mid i\in[n],\,j\in[\alpha]\}$, of size $n\alpha$ \\
		\midrule
		\multicolumn{2}{l}{\textbf{Signal Flow Graph and Linking Conditions}} \\
		\midrule
		$\sV_t$ & Set of vertices at stage $t$ \\
		$\mathcal{G}_t$ & Set of global encoding vectors at stage $t$ \\
		$\mathscr{M}_t$ & Gammoid with ground set $\mathcal{V}_t\cup\mathcal{V}_{t+1}$ \\
		$\mathscr{C}_t$ & Independent sets of $\mathscr{M}_t$ \\
		$(s,t)$ & $F$-pair: adjacent identical failure pair of stages ($s<t$) \\
		$\sF{a,b}$ & Set of failed nodes from stage $a$ to $b$ \\
		$\sP{a,b}(i)$ & Set of choice functions of node $i$ from stage $a$ to $b$ \\
		$\mV{s}$ & Active vertices at stage $s$\\
		$\mathcal{D}_s$ & Failed vertices at stage $s$\\
		$\mathbf{N}_s$ & Available nodes at stage $s$\\
		\midrule
		\multicolumn{2}{l}{\textbf{Linear Representation of Gammoids (Section \ref{sec: linear})}} \\
		\midrule
		$(i,j)_s$ & Stage-indexed index corresponding to vertex $v_s(i,j)$ \\
		$\mathcal{I},\,\mathcal{I}_t$ & Index set and its restriction to stages $t$ \\
		$\mathcal{V}(\mathcal{I})$ & Vertices corresponding to $\mathcal{I}$ \\
		$\mathcal{G}(\mathcal{I})$ & Global encoding vectors corresponding to $\mathcal{I}$ \\
		$\mathbf{L}_t$ & Matrix of local encoding coefficients at stage $t$ \\
		\bottomrule
	\end{tabular}
\end{table}

When a node fails, we replace the failed node by a new node, called the {\em newcomer}, so that the $(n,k)$ recovery property is maintained after the repair. The surviving nodes that participate in the repair of the failed node are called the {\em helpers}. The content stored in the newcomer is computed by downloading $\beta$ packets from each of the $d=n-1$ surviving nodes. The {\em repair bandwidth} is defined as the total number of packet transmissions from the helpers to the newcomer during the repair process, and is therefore equal to $d\beta$. In this paper, we consider the repair of a single failed node, and assume that all remaining $n-1$ nodes participate in the repair process. Thus, the {\em repair degree} is $d=n-1$. We consider the {\em functional-repair} mode of regeneration, in which the content stored in the newcomer need not be identical to the content stored in the failed node. Instead, it is sufficient that the $(n,k)$ recovery property be maintained after the repair process.

For any fixed file size $B$, the inequality in~\eqref{eq:tradeoff} establishes a fundamental tradeoff between the storage per node and the repair bandwidth. To ensure successful file recovery, any collection of $k$ nodes must collectively store at least $B$ packets. Therefore, each node must store at least $B/k$ packets. One extreme point of the tradeoff is obtained by minimizing the storage per node, namely by setting $\alpha=B/k$. In this regime, the minimum repair bandwidth is found by identifying the smallest value of $\beta$ such that $(d-i+1)\beta$ is at least $B/k$ for all $i=1,2,\ldots,k$. The minimum value of $\beta$ is  $B/(k(d-k+1))$, which gives the {\em minimum-storage regenerating} (MSR) point
\[
(\alpha_\mathrm{MSR}, \beta_\mathrm{MSR}) :=
\left(\frac{B}{k}, \frac{B}{k(d-k+1)} \right).
\]

The other extreme point is obtained by minimizing the repair bandwidth. Ignoring the constraint imposed by $\alpha$ in~\eqref{eq:tradeoff}, we first find the minimum value of $\beta$ satisfying
\[
B \leq \sum_{i=1}^k (d-i+1)\beta.
\]
The minimum value of $\beta$ in this case is
\[
\frac{2B}{k(2d-k+1)}.
\]
For this choice of $\beta$, the smallest value of $\alpha$ that does not violate~\eqref{eq:tradeoff} is $\alpha=d\beta$, namely
\[
\alpha=\frac{2Bd}{k(2d-k+1)}.
\]
The corresponding operating point is called the {\em minimum-bandwidth regenerating} (MBR) point:
\[
(\alpha_\mathrm{MBR}, \beta_\mathrm{MBR}) :=
\left(\frac{2Bd}{k(2d-k+1)}, \frac{2B}{k(2d-k+1)} \right).
\]

Other points on the tradeoff curve can be obtained by choosing $\alpha$ between $B/k$ and $2Bd/(k(2d-k+1))$, and then solving for the minimum value of $\beta$ that satisfies~\eqref{eq:tradeoff}. The tradeoff curve is piecewise linear. Its vertices are given by
\begin{equation}
	(\alpha,\beta) =
        \frac{2B}{2k(d-\ell+1) - (k-\ell)(k-\ell+1)}
        (d-\ell+1,1),
	\label{eq:op}
\end{equation}
for $\ell=1,2,\ldots,k$. When $\ell=k$, the vertex corresponds to the MSR point, and when $\ell=1$, it corresponds to the MBR point.

\begin{figure}
	\begin{center}
	\includegraphics[width=9cm]{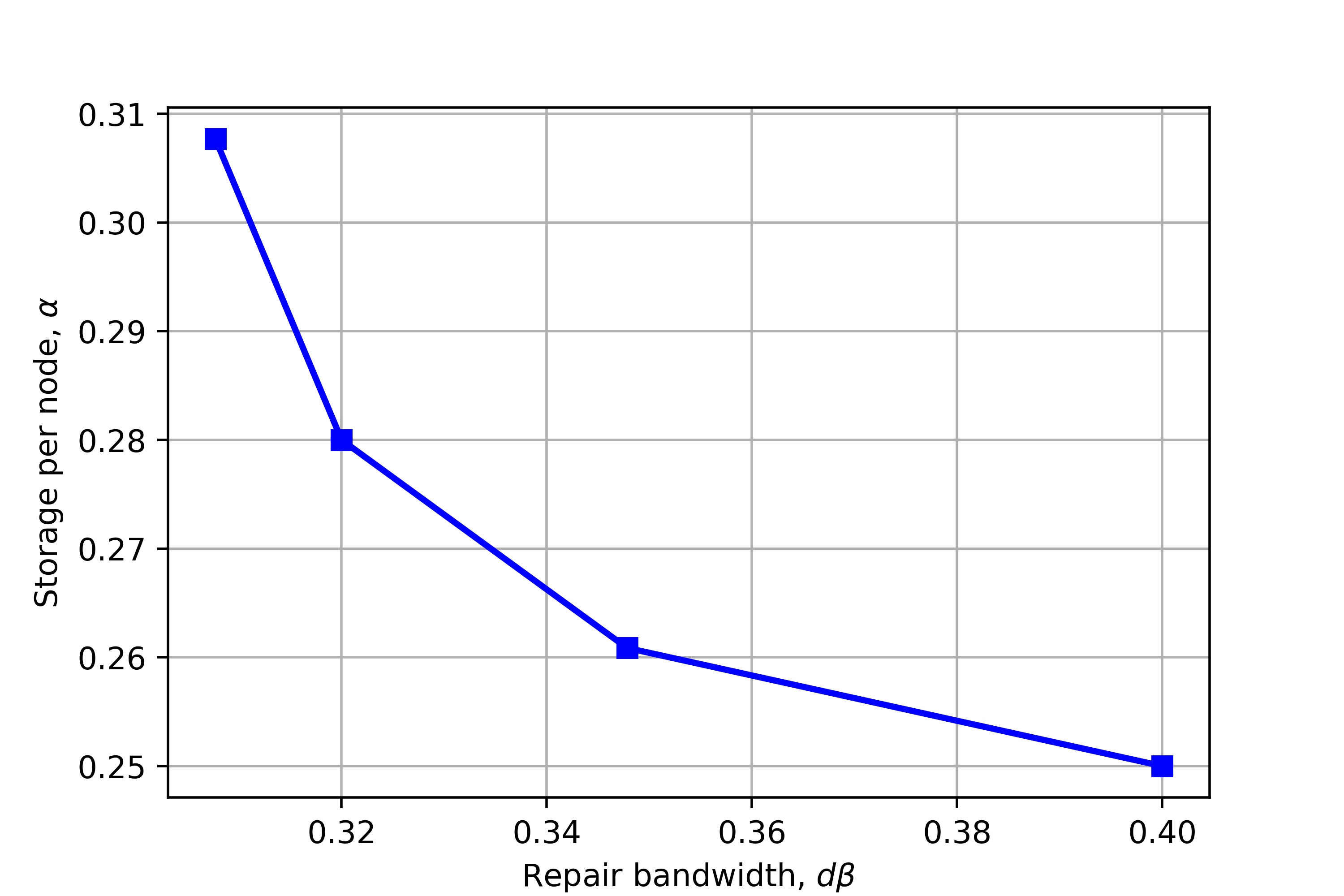}
	\end{center}
\caption{Normalized tradeoff curve between storage and repair bandwidth for the distributed storage system with parameters $B=1$, $k=4$, and $d=8$.}
\label{fig:tradeoff}
\end{figure}

An example of a tradeoff curve with four vertex points is shown in Fig.~\ref{fig:tradeoff}. The figure plots the normalized storage per node, $\alpha/B$, against the normalized repair bandwidth, $d\beta/B$. For example, the MSR point, located at the lower-right corner of the curve, can be achieved by choosing $\alpha=5$ and $\beta=1$. In this example, the file size is $B=20$. Thus, each node stores one quarter of the total file size. The normalized repair bandwidth is
\[
\frac{d\beta}{B} = \frac{8}{20} = 40\%.
\]
The normalized repair bandwidth can be further reduced by choosing the parameters $B=23$, $\alpha=6$, and $\beta=1$. With this choice, the normalized storage per node is
\[
\frac{\alpha}{B} = \frac{6}{23} \approx 0.261,
\]
whereas the normalized repair bandwidth is
\[
\frac{d\beta}{B} = \frac{8}{23} \approx 0.348.
\]
Thus, each node stores slightly more than one quarter of the file, while the repair bandwidth is reduced to approximately $34.8\%$ of the file size.

The other two vertices on the tradeoff curve are $(8/25,7/25)$ and $(4/13,4/13)$ in normalized coordinates. The last point, $(4/13,4/13)$, is the MBR point. At this point, repairing a single node failure requires a repair bandwidth equal to approximately $30.8\%$ of the total file size.

In addition to minimizing the repair bandwidth, we also aim to reduce the disk I/O cost during node repair. Ideally, each helper reads exactly $\beta$ packets and transfers these packets directly to the newcomer without performing any arithmetic operation. Such a repair scheme is help-by-transfer and access-optimal, since the number of packets read by each helper is equal to the number of packets it transmits. This minimizes both the disk access and the computational workload at the helper nodes. The newcomer then collects the packets from all helpers, computes the required linear combinations, and stores $\alpha$ new packets. The main objective of this paper is to construct regenerating codes that achieve all vertex points on the fundamental tradeoff curve with help-by-transfer and access-optimal repair, under the assumption that the repair degree is $d=n-1$.

\section{Linear Regenerating Code as a Discrete-time Dynamical System}

\label{sec:linear_regenerating_code}

We implement help-by-transfer functional repair using linear regenerating codes. In the remainder of this paper, we assume that $d=n-1$, i.e., all surviving nodes participate in the repair of a failed node. By suitably normalizing the parameters $B$, $\alpha$, and $\beta$, we may choose the file size $B$ such that the variable $\beta$ in~\eqref{eq:op} is equal to one. Thus, we consider the parameters
\begin{align}
	\beta &= 1, \nonumber \\
	\alpha &= d-\ell+1 = n-\ell, \nonumber\\
	B &= k\alpha - (1+2+\cdots+(k-\ell)), \label{eq: para}
\end{align}
where $\ell$ is an integer in $\{1,2,\ldots,k\}$. When $\ell=1$, the corresponding operating point is the minimum-bandwidth regenerating point, and when $\ell=k$, it is the minimum-storage regenerating point. It is straightforward to verify that, for each $\ell$, the parameters in~\eqref{eq: para} achieve equality in the fundamental tradeoff bound in~\eqref{eq:tradeoff}. Throughout this paper, we fix an arbitrary $\ell$ and present the construction for the corresponding set of parameters.

Under these parameter settings, the number of source packets is always at least $n-1$.

\begin{lemma} \label{lemma:B_larger_than_n_minus_1}
If $n\geq k+1$, then
\[
    B \geq n-1
\]
under the parameter setting in~\eqref{eq: para}.
\end{lemma}

\begin{proof}
Expanding the expression of $B$ in~\eqref{eq: para}, we have
\begin{align*}
    B
    &= k(n-\ell)-\frac{(k-\ell)(k-\ell+1)}{2}\\
    &= \frac{1}{2}\bigl(2kn-2k\ell-(k^2-2k\ell+\ell^2+k-\ell)\bigr)\\
    &= \frac{1}{2}\bigl(2kn-k^2-\ell^2-k+\ell\bigr).
\end{align*}
Since $n\geq k+1$, it follows that
\begin{align*}
B-(n-1)
&= \frac{1}{2}\bigl(2(k-1)n-k^2-\ell^2-k+\ell+2\bigr)\\
&\geq \frac{1}{2}\bigl(2(k-1)(k+1)-k^2-\ell^2-k+\ell+2\bigr)\\
&= \frac{1}{2}\bigl(k^2-\ell^2-(k-\ell)\bigr)\\
&= \frac{1}{2}(k-\ell)(k+\ell-1)\geq 0.
\end{align*}
This proves that $B\geq n-1$.
\end{proof}

In the framework of linear regenerating codes, each packet is regarded as an element of a finite field $\FF{q}$ of size $q$, where $q$ is a prime power. All encoding and decoding operations are linear transformations over $\FF{q}$. The data file to be stored in the distributed storage system is divided into chunks, each containing
\[
    B=k\alpha-(1+2+\cdots+(k-\ell))
\]
finite-field symbols. Each storage node stores $\alpha=n-\ell$ symbols. To achieve optimal disk I/O cost, each helper node reads exactly one finite-field symbol and sends it to the newcomer. Since each helper only transmits one stored symbol to the newcomer, no computation is required on the helper side.

Unlike the construction of regenerating codes in~\cite{Wu09}, the operations in our construction are time-dependent. Thus, the system is time-inhomogeneous. In our formulation, time is discrete and is referred to as a {\em stage}. Stage $0$ denotes the time immediately after the initialization of the storage system, and the stage index is incremented by one after each single-node failure and recovery. We describe the proposed regenerating code as a discrete-time linear dynamical system.

Let the $B$ source packets be denoted by $Q_1,Q_2,\ldots,Q_B$. We represent the data file as a row vector\footnote{All vectors in this paper are row vectors by default.}
\[
    \mathbf{m} :=(Q_1,Q_2,\ldots,Q_B)
\]
in a $B$-dimensional vector space over $\FF{q}$. A coded symbol stored in a storage node is a linear combination of the source packets, and hence can be represented as a dot product
\[
    \mathbf{m}\cdot \mathbf{v},
\]
where $\mathbf{v}$ is called the associated {\em global encoding vector}. Mathematically, the global encoding vector is an element of the dual space and has the same dimension as the message vector $\mathbf{m}$. The content of a storage node is represented by $\alpha$ global encoding vectors in $\FF{q}^B$.

For $i\in\{1,2,\ldots,n\}$, $j\in\{1,2,\ldots,\alpha\}$, and $t\geq 0$, let $S_t(i,j)$ denote the $j$-th finite-field symbol stored in the $i$-th storage node at stage $t$. We also refer to $S_t(i,j)$ as the $j$-th packet in node $i$ at stage $t$. The global encoding vector of $S_t(i,j)$ is denoted by $\mathbf{e}_t(i,j)$, so that
\[
    S_t(i,j)=\mathbf{m}\cdot \mathbf{e}_t(i,j),
\]
for all $i\in\{1,2,\ldots,n\}$, $j\in\{1,2,\ldots,\alpha\}$, and $t\geq 0$.

For a fixed stage $t$ and a fixed node index $i$, the vectors $\mathbf{e}_t(i,j)$, for $j=1,2,\ldots,\alpha$, represent the {\em state} of storage node $i$ at stage $t$. The linear dynamical system describes how these states evolve over time.

\medskip

\noindent{\bf Data recovery for a data collector.}
In order to maintain the $(n,k)$ recovery property, for any $t\geq 0$ and for any $k$ distinct node indices $i_1,i_2,\ldots,i_k$, with
\[
    \mathbf{K}:=\{i_1,i_2,\ldots,i_k\}\subseteq \{1,2,\ldots,n\},
\]
we require the global encoding vectors
\[
    \mathbf{e}_t(i,j), \quad
    i\in \mathbf{K}, \quad j=1,2,\ldots,\alpha,
\]
to span the entire vector space $\FF{q}^B$. This requirement ensures that a data collector can recover the original data file from any $k$ storage nodes, regardless of the values of the source symbols $Q_1,\ldots,Q_B$. Moreover, we require the $(n,k)$ recovery property to be maintained for an unbounded period of time, in order to guarantee long-term data reliability.

\medskip

\noindent{\bf Repair of failed nodes.}
For stage $t\geq 0$, let $F_t$ denote the index of the failed node, where $F_t$ may be any element of $\{1,2,\ldots,n\}$. Node $F_t$ is repaired during the transition from stage $t$ to stage $t+1$. Every node $i\neq F_t$ reads one finite-field symbol stored in its memory and transmits it to the newcomer. We denote by $p_t(i)$ the index of the packet sent from node $i$ to the newcomer at stage $t$, where
\[
    p_t(i)\in \{1,2,\ldots,\alpha\}.
\]
We call $p_t(i)$ the {\em choice function}. The value of $p_t(F_t)$ is undefined.

After receiving the repair data from the surviving nodes, the newcomer regenerates its content by computing
\begin{equation}
    S_{t+1}(F_t,j)
    =
    \sum_{i\neq F_t} b_t(i,j) S_t(i,p_t(i)),     
    \quad j=1,2,\ldots,\alpha,
    \label{eq:local_repair}
\end{equation}
where the summation is over all indices
$i\in\{1,2,\ldots,n\}\setminus\{F_t\}$. The coefficients $b_t(i,j)$ take values in $\FF{q}$ and are called the {\em local encoding coefficients}. The global encoding vector associated with $S_{t+1}(F_t,j)$ is obtained from the state-transition equation
\begin{equation}
	\mathbf{e}_{t+1}(F_t,j)
    =
    \sum_{i\neq F_t} b_t(i,j)\mathbf{e}_t(i,p_t(i)),
	\label{eq:transfer1}
\end{equation}
for $j=1,2,\ldots,\alpha$.

Meanwhile, for each $i\neq F_t$, the content of node $i$ remains unchanged during the transition from stage $t$ to stage $t+1$. Hence,
\begin{equation}
	S_{t+1}(i,j)=S_t(i,j)
    \quad\text{and}\quad
	\mathbf{e}_{t+1}(i,j)=\mathbf{e}_t(i,j),
	\label{eq:transfer2}
\end{equation}
for $i\neq F_t$ and $j=1,2,\ldots,\alpha$.

Equations~\eqref{eq:transfer1} and~\eqref{eq:transfer2} mathematically capture the help-by-transfer property: a failed node is repaired by having each helper read and transmit one stored symbol, without performing any computation. If the code parameters are chosen at a vertex point of the tradeoff curve in~\eqref{eq:tradeoff}, and if the local encoding coefficients and the choice function can be selected so that the $(n,k)$ recovery property is maintained for an unbounded period of time, then both the repair bandwidth and the disk I/O cost are optimal.

\medskip

\noindent{\bf Definition of the linear regenerating code.}
An $(n,k,d=n-1,\alpha,\beta=1,B)$ linear functional-repair regenerating code over $\FF{q}$ is specified as follows. The system state at stage $t$, for $t=0,1,2,\ldots$, consists of the global encoding vectors
\[
    \mathbf{e}_t(i,j)\in \FF{q}^{B},
    \quad
    i=1,2,\ldots,n,\quad j=1,2,\ldots,\alpha.
\]
For a message vector $\mathbf{m}\in \FF{q}^{B}$, the $j$-th packet stored in node $i$ at stage $t$ is
\[
    S_t(i,j)=\mathbf{m}\cdot \mathbf{e}_t(i,j).
\]
The initial state is given by a collection of vectors
\[
    \{\mathbf{e}_0(i,j): i=1,2,\ldots,n,\ j=1,2,\ldots,\alpha\}.
\]

At each stage $t$, one node $F_t\in [n]$ may fail. The repair rule consists of a choice function
\[
    p_t:[n]\setminus\{F_t\}\longrightarrow [\alpha]
\]
and local encoding coefficients
\[
    b_t(i,j)\in \FF{q},
    \quad i\in [n]\setminus\{F_t\},\quad j=1,2,\ldots,\alpha.
\]
Each surviving node $i\neq F_t$ reads and transmits the single stored packet $S_t(i,p_t(i))$
to the newcomer. The newcomer then computes its new stored packets according to equation~\eqref{eq:local_repair}.
Equivalently, the global encoding vectors evolve according to~\eqref{eq:transfer1} and \eqref{eq:transfer2}.

We want to determine the local encoding coefficients $b_t(i,j)$ and the choice function $p_t(i)$ for an unbounded sequence of repairs so that the $(n,k)$ recovery holds at stage $0$ and continues to hold after every subsequent update specified by~\eqref{eq:transfer1} and~\eqref{eq:transfer2}.

\medskip

A few important remarks are in order.

\renewcommand{\labelenumi}{(\roman{enumi}).}
\begin{enumerate}
\item Maintaining the $(n,k)$ recovery property is more straightforward for exact-repair regenerating codes, because the system does not involve the notion of time evolution: the content of the newcomer is identical to that of the failed node. Hence, if the $(n,k)$ recovery property holds before a node failure, then it also holds after the repair. The design of exact-repair regenerating codes can therefore be reduced to a purely algebraic problem. However, as shown in~\cite{Tian2014}, there can be a non-vanishing gap between the optimal storage-bandwidth tradeoff for functional repair and that for exact repair. In this paper, we focus on functional repair and aim to achieve the points on the functional-repair storage-bandwidth tradeoff curve.

\item Since the stage index may increase indefinitely, the information flow graph that models the system is, in principle, an infinite graph. Therefore, we cannot directly apply the functional-repair regenerating codes described in~\cite{DGWR2010}, because the required lower bound on the finite-field size grows with the size of the information flow graph. Moreover, random network coding is not suitable for our purpose, since we require the data collector to recover the original data file with probability one after every repair. Thus, all operations in this paper are deterministic.

\item We cannot control or predict which node fails at stage $t$. Our goal is to maintain the $(n,k)$ recovery property indefinitely, for every possible failure pattern. The infinite sequence of node failures
\[
    F_0,F_1,F_2,\ldots
\]
may be regarded as being generated by an external mechanism. For any realization of this sequence, the local encoding coefficients should be chosen so that the system continues to satisfy the recovery property.

The choice function, however, must be causal: the packet selected by a helper at stage $t$ may depend only on the failure history available up to stage $t$, and not on future failures. For practical implementability, we further require this dependence to have finite memory. That is, the value of $p_t(i)$ depends only on a maintained state whose size is bounded
independently of $t$. In this sense, $p_t(i)$ is a {\em causal finite-memory} function.

During the repair of node $F_t$, both the choice function $p_t(i)$ and the local encoding coefficients $b_t(i,j)$ may depend on the available finite-memory state. For notational simplicity, this dependence is suppressed.
\end{enumerate}

\section{Prerequisites on Matroids and Gammoids}
\label{sec: matroid}

We use matroid theory in the construction and analysis of the proposed scheme. Matroids provide an abstract framework for studying independence, generalizing both linear independence in vector spaces and acyclicity in graphs~\cite{Welsh76,Oxley}. This abstraction is particularly useful in our setting, because the data recovery condition can be naturally expressed as a spanning condition on collections of global encoding vectors.

\subsection{Matroids}

A matroid $\mathscr{M}=(\mathcal{X},\mathscr{C})$ consists of a finite set $\mathcal{X}$, called the {\em ground set}, and a collection $\mathscr{C}$ of subsets of $\mathcal{X}$ satisfying the following properties:

\begin{enumerate}
\item[$I$1.] $\emptyset \in \mathscr{C}$;
\item[$I$2.] if $\mathcal{B}\in\mathscr{C}$ and $\mathcal{A}\subseteq \mathcal{B}$, then $\mathcal{A} \in \mathscr{C}$;
\item[$I$3.] if $\mathcal{A}$, $\mathcal{B}\in\mathscr{C}$ and $|\mathcal{A}|< |\mathcal{B}|$, then we can find an element $x$ in $\mathcal{B} \setminus \mathcal{A}$ such that $\mathcal{A} \cup \{x\}$ is in $\mathscr{C}$.
\end{enumerate}
The sets in $\mathscr{C}$ are called the {\em independent sets} of the matroid. A subset of $\mathcal{X}$ that is not in $\mathscr{C}$ is called dependent.

The three properties above abstract the basic properties of linear independence in vector spaces. Equivalently, a matroid can be characterized by a rank function. A matroid $\mathscr{M}=(\mathcal{X},\rho)$ consists of a finite set $\mathcal{X}$ and an integer-valued function $\rho:2^{\mathcal{X}}\rightarrow \mathbb{Z}_{\geq 0}$ satisfying the following properties for all $\mathcal{A},\mathcal{B}\subseteq \mathcal{X}$:

\begin{enumerate}
\item[$R1$.] $ 0 \leq \rho(\mathcal{A}) \leq |\mathcal{A}|$;

\item[$R2$.] $\rho(\mathcal{A})\leq \rho(\mathcal{B})$ for $\mathcal{A} \subseteq \mathcal{B} \subseteq \mathcal{X}$;

\item[$R3$.] $\rho(\mathcal{A})+\rho(\mathcal{B}) \geq
\rho(\mathcal{A}\cup \mathcal{B})+\rho(\mathcal{A}\cap\mathcal{B}) $.
\end{enumerate}
Given the independent-set formulation, the rank of a subset $\mathcal{A}\subseteq \mathcal{X}$ is
\[
    \rho(\mathcal{A})
    =
    \max\{|\mathcal{I}|:\mathcal{I}\subseteq \mathcal{A},\ \mathcal{I}\in\mathscr{C}\}.
\]
Conversely, given a rank function $\rho$, a subset $\mathcal{A}\subseteq \mathcal{X}$ is independent if and only if $\rho(\mathcal{A})=|\mathcal{A}|$.

\begin{example}[Vector matroid]
Given a matrix over a field $\mathbb{F}$, let the ground set $\mathcal{X}$ be the set of column labels of the matrix. A subset $\mathcal{A}\subseteq \mathcal{X}$ is defined to be {\em independent} if the columns labeled by the elements of $\mathcal{A}$ are linearly independent over $\mathbb{F}$. This defines a matroid, called the {\em vector matroid} associated with the matrix.
\end{example}

\begin{remark}
In a vector matroid, the ground set consists of column labels rather than the column vectors themselves. Thus, two distinct elements of the ground set may correspond to identical column vectors. In the remainder of this paper, when we refer to the vector matroid formed by a collection of vectors, the ground set is taken to be the index set of these vectors, and a subset is independent if and only if the corresponding vectors are linearly independent.
\end{remark}

We also recall three standard notions in matroid theory.

\begin{itemize}
\item Two matroids $\mathscr{M}=(\mathcal{X},\mathscr{C})$ and $\mathscr{M}'=(\mathcal{X}',\mathscr{C}')$ are said to be {\em isomorphic} if there exists a bijection $\theta:\mathcal{X}\rightarrow \mathcal{X}'$ such that, for every subset $\mathcal{A}\subseteq \mathcal{X}$, the set $\mathcal{A}$ is independent in $\mathscr{M}$ if and only if $\theta(\mathcal{A})$ is independent in $\mathscr{M}'$, where
\[
    \theta(\mathcal{A}) :=\{\theta(x):x\in\mathcal{A}\}.
\]

\item A matroid is said to be {\em linearly representable} over a field $\mathbb{F}$ if it is isomorphic to a vector matroid over $\mathbb{F}$. It is said to be linearly representable if it is linearly representable over some field.

\item Given a matroid $\mathscr{M}=(\mathcal{X},\mathscr{C})$ and a subset $\mathcal{A}\subseteq \mathcal{X}$, the {\em restriction} of $\mathscr{M}$ to $\mathcal{A}$ is the matroid
\[
    \mathscr{M}|_{\mathcal{A}}
    =
    (\mathcal{A},\mathscr{C}|_{\mathcal{A}}),
\]
where
\[
    \mathscr{C}|_{\mathcal{A}}
    =
    \{\mathcal{I}\subseteq \mathcal{A}:\mathcal{I}\in \mathscr{C}\}.
\]
\end{itemize}

\subsection{Linkings and Gammoids}
\label{sec:linking_gammoid}
A {\em gammoid} is a matroid defined from a directed graph, where independence is characterized by the existence of vertex-disjoint paths, called {\em linkings}. We first define the notion of linking.

Let $G=(\sV,\sE)$ be a directed graph, where $\sV$ is the vertex set and $\sE\subseteq \sV\times \sV$ is the edge set. A {\em path} in $G$ is a sequence
\[
    P=(v_0,v_1,\ldots,v_r)
\]
of distinct vertices of $G$, with $r\geq 0$, such that $(v_{i-1},v_i)\in\sE$ for all $i=1,2,\ldots,r$. The first vertex $v_0$ is called the {\em initial vertex} of $P$, and the last vertex $v_r$ is called the {\em terminal vertex} of $P$. The case $r=0$ is allowed, in which case the path consists of a single vertex. Two paths are said to be vertex-disjoint if their vertex sets are disjoint.

Let $\mathcal{A}$ and $\mathcal{B}$ be two subsets of $\mathcal{V}$. We say that there exists a {\em linking of $\mathcal{A}$ into $\mathcal{B}$} if there exist an injective map $f:\mathcal{A}\rightarrow \mathcal{B}$ and a collection of mutually vertex-disjoint paths
\[
    \{P_a:a\in\mathcal{A}\}
\]
such that, for each $a\in\mathcal{A}$, the path $P_a$ has initial vertex $a$ and terminal vertex $f(a)$. If, in addition, $f$ is a bijection from $\mathcal{A}$ onto $\mathcal{B}$, then we say that $\mathcal{A}$ can be {\em linked onto} $\mathcal{B}$.

For a directed graph $G=(\sV,\sE)$ and a fixed subset $\sB\subseteq \sV$, let $L(G,\sB)$ denote the collection of subsets of $\mathcal{V}$ that can be linked into $\sB$. It is known that $L(G,\sB)$ is the collection of independent sets of a matroid on the ground set $\mathcal{V}$~\cite[Chapter 13]{Welsh76}. This matroid is called a {\em strict gammoid}.

\begin{definition} A {\em gammoid} is defined as the restriction of a strict gammoid to some subset of its ground set.
\end{definition}

The notion of a gammoid is particularly relevant in this work, as in network coding more broadly, because finite gammoids admit linear representations over sufficiently large finite fields.

\begin{theorem}[\cite{Mason72,Lindstrom73}]
For every finite gammoid, there exists a positive integer $q_0$ such that the gammoid is linearly representable over every finite field $\FF{q}$ with $q\ge q_0$.
\end{theorem}

\begin{example}
 Consider the graph $G$ in Fig \ref{fig:gammoid}. Let $\sB=\{v_1,v_4\}$. The strict gammoid $L(G,\sB)$ has ground set $\{v_1,v_2,v_3,v_4\}$ and the collection of all independent sets $$\{\emptyset, \{v_1\},\{v_2\},\{v_3\},\{v_4\},\{v_1,v_2\}, \{v_1,v_3\},\{v_1,v_4\}\}.$$
As an example, the set $\{v_1,v_2\}$ is independent, since it can be linked onto $\mathcal{B}$ by the paths $P_1=(v_1)$ and $P_2=(v_2,v_3,v_4)$.

Furthermore, the gammoid obtained by restricting $L(G,\sB)$ to $\sA=\{v_1,v_3\}\subseteq\mathcal{V}$ has ground set $\{v_1,v_3\}$ and the collection of independent sets $\{\emptyset, \{v_1\},\{v_3\},\{v_1,v_3\}\}$. 
  \label{ex:gammoid}
\end{example}

\begin{figure}[ht]
    \centering
\begin{tikzpicture}
  \node (v1) [circle ,fill=black,inner sep=1.5pt] at (0,0) { };
  \node (t1) [circle] at (0,-0.5) {$v_1$};
  \node (v2) [circle ,fill=black,inner sep=1.5pt] at (1.5,0) { };
  \node (t2) [circle] at (1.5,-0.5) {$v_2$};
  \node (v3) [circle ,fill=black,inner sep=1.5pt] at (3,0) { };
  \node (t3) [circle] at (3,-0.5) {$v_3$};
  \node (v4) [circle ,fill=black,inner sep=1.5pt] at (4.5,0) { };
  \node (t4) [circle] at (4.5,-0.5) {$v_4$};
  \draw[postaction={decorate, decoration={markings,
      mark=at position 0.5 with {\arrow{>}}}}] (v1) -- (v2);
  \draw[postaction={decorate, decoration={markings,
      mark=at position 0.5 with {\arrow{>}}}}] (v2) -- (v3);
  \draw[postaction={decorate, decoration={markings,
      mark=at position 0.5 with {\arrow{>}}}}] (v3) -- (v4);
\end{tikzpicture}
  \caption{The directed graph that defines the gammoid in Example~\ref{ex:gammoid}.}
    \label{fig:gammoid}
\end{figure}

\section{Signal Flow Graph Representation} \label{sec: graph}

We use a directed graph to visualize the dependencies among the encoded packets. We call this directed graph a {\em signal flow graph}.

We abstract the discrete-time linear dynamical system described in Section~\ref{sec:linear_regenerating_code} by a signal flow graph. The vertices of the signal flow graph are partitioned into stages. For a positive integer $t$, stage $t$ corresponds to the state of the storage system after the $t$-th repair, namely after the repair of the failed nodes $F_0,F_1,\ldots,F_{t-1}$. Stage $0$ represents the initial state of the storage system. Each vertex in the graph represents a corresponding global encoding vector.

An edge from a vertex $v$ to a vertex $w$ in the signal flow graph signifies that the global encoding vector associated with $w$ depends linearly on the global encoding vector associated with $v$. The edge set is determined by the choice functions. Each edge is labeled by an element of $\FF{q}$, representing the corresponding local encoding coefficient.

Specifically, suppose that the in-neighbors of a vertex $w$ are $v_1,\ldots,v_m$, and that the edge from $v_r$ to $w$ is labeled by $b(v_r,w)\in\FF{q}$. Then the global encoding vector corresponding to $w$ satisfies
$$
    \be_w=\sum_{r=1}^m b(v_r,w)\be_{v_r}.
$$
Thus, the edges of the signal flow graph represent the linear dependencies among the global encoding vectors.

Formally, we construct the signal flow graph as follows. Let
\begin{equation}
 \sJ := \{(i,j):\, i=1,2,\ldots, n, \ j=1,2,\ldots, \alpha\}
 \label{eqn:indexset}
\end{equation}
be an index set consisting of $n\alpha$ indices. For each nonnegative integer $t$, there are $n\alpha$ vertices at stage $t$, labeled by 
$v_t(i,j)$ for $(i,j) \in \sJ$. The vertex $v_t(i,j)$ corresponds to the global encoding vector $\mathbf{e}_t(i,j)$ at stage $t$.

In addition, we add an auxiliary stage $-1$ to represent the initialization of the distributed storage system. At stage $-1$, we create $B$ vertices and label them by $1,2,\ldots,B$. For $b=1,2,\ldots,B$, vertex $b$ at stage $-1$ represents the global encoding vector of the source packet $Q_b$, which is the $b$-th unit vector in the standard basis of $\FF{q}^B$:
$$
\mathbf{u}_b := (\underbrace{0,0,\ldots, 0}_{b-1 \text{ zeros}}, 1, \underbrace{0,0,\ldots 0}_{B-b \text{ zeros}}).
$$

All edges in the signal flow graph connect vertices of consecutive stages and are labeled by the local encoding coefficients.

At stage $0$, for each $b=1,2,\ldots,B$ and each $(i,j)\in\sJ$, we draw a directed edge from vertex $b$ at stage $-1$ to vertex $v_0(i,j)$ at stage $0$. The label of this edge is the coefficient of $\mathbf{u}_b$ in the expansion of $\mathbf{e}_0(i,j)$ with respect to the standard basis $\{\mathbf{u}_1,\ldots,\mathbf{u}_B\}$. That is,
\[
    \be_0(i,j)=\sum_{b=1}^B [\be_0(i,j)]_b \bu_b,
\]
where $[\be_0(i,j)]_b$ denotes the $b$-th coordinate of $\be_0(i,j)$. In this way, the vertices $v_0(i,j)$ represent the initial global encoding vectors $\mathbf{e}_0(i,j)$.

For each nonnegative integer $t$ and each $(i,j)\in\sJ$ such that node $i$ is not the failed node at stage $t$, i.e., $i\neq F_t$, we draw a directed edge from vertex $v_t(i,j)$ to vertex $v_{t+1}(i,j)$ and label it by the identity element $1\in\FF{q}$. These edges represent the fact that, for each storage node that does not fail at stage $t$, the stored content remains unchanged at stage $t+1$. We refer to such edges as {\em carry-forward edges}.

For each index $(F_t,j)$, where $t\geq 0$ and $j=1,2,\ldots,\alpha$, we draw $d=n-1$ directed edges from stage $t$ to stage $t+1$, all terminating at the vertex $v_{t+1}(F_t,j)$. These edges emanate from the vertices
\[
    v_t(i,p_t(i)), \qquad i\in\{1,2,\ldots,n\}\setminus\{F_t\},
\]
and are labeled by the corresponding local encoding coefficients $b_t(i,j)$. This represents the state-transition relation in~\eqref{eq:transfer1}. We refer to such edges as {\em repair edges}.

An example of a signal flow graph is shown in Fig.~\ref{fig:graph}. In this example, there are $n=4$ storage nodes. Each node stores $\alpha=2$ packets, and the repair bandwidth per helper is $\beta=1$. According to~\eqref{eq:tradeoff}, these parameters support a file size of $B=4$. Hence, we place four vertices at stage $-1$, representing the four source symbols. The gray edges from stage $-1$ to stage $0$ represent the initialization of the storage system. At each stage, each node contains two vertices, representing its two global encoding vectors.

The red ``X'' marks the failed node at each stage. For instance, node $2$ fails at stage $0$, and therefore there is no outgoing edge from node $2$ at stage $0$. Each vertex in node $2$ at stage $1$ is connected to one vertex in each surviving node at stage $0$. The choice function at stage $0$ is
\[
    p_0(1)=1,\qquad p_0(3)=1,\qquad p_0(4)=2.
\]
Hence, the vertices in node $2$ at stage $1$ are connected to the first vertex in node $1$, the first vertex in node $3$, and the second vertex in node $4$ at the previous stage. These connections are colored in blue. If a node does not fail, its packets are carried forward, as represented by the black edges. The labels on the edges are omitted for clarity.

In the realization depicted in Fig.~\ref{fig:graph}, node $3$ fails at stage $1$, and node $4$ fails at stage $2$. The $(4,2)$ recovery property requires that, if any two nodes are selected at any stage, then the four corresponding vertices should be linked to the four source vertices at stage $-1$ by four vertex-disjoint paths. We formally establish this connection in the next section.

\begin{figure}[ht]
	\centering
	\begin{tikzpicture}[
    node/.style={draw, circle, fill=black, minimum size=2pt, inner sep=0pt},
    blueNode/.style={draw=blue, circle, fill=blue, minimum size=2pt, inner sep=0pt},
    stage/.style={font=\small\bfseries, align=center},
    F/.style={font=\small, align=center},
    group/.style={},
    dashline/.style={gray, dashed, line width=0.5pt},
    redX/.style={red, line width=1pt}
]

\def\regularSpacing{2.5}
\def\wideSpacing{3.75} 

\def\groupSpacing{1.6}  
\def\pairOffset{0.2}  
\def\topY{3*\groupSpacing}

\pgfmathsetmacro{\stageZeroX}{\regularSpacing}
\pgfmathsetmacro{\stageOneX}{\regularSpacing+\wideSpacing}
\pgfmathsetmacro{\stageTwoX}{\regularSpacing+\wideSpacing+\regularSpacing}
\pgfmathsetmacro{\stageThreeX}{\regularSpacing+\wideSpacing+2*\regularSpacing}

\node[stage] at (0, -1.2) (stageLabelM1) {Stage -1};
\foreach \y in {0,1,2,3} {
    \node[node] at (0, \topY-\y*\groupSpacing) (m1-\y) {};
}

\node[stage] at (\stageZeroX, -1.2) (stageLabel0) {Stage 0};
\node [F] at (\stageZeroX, -1.6) {$F_0=2$};
\foreach \group in {0,1,2,3} {
    \node[node] at (\stageZeroX, \topY-\group*\groupSpacing+\pairOffset) (s0-\group-a) {};
    \node[node] at (\stageZeroX, \topY-\group*\groupSpacing-\pairOffset) (s0-\group-b) {};
}

\node[stage] at (\stageOneX, -1.2) (stageLabel1) {Stage 1};
\node [F] at (\stageOneX, -1.6) {$F_1=3$};
\foreach \group in {0,1,2,3} {
    \node[node] at (\stageOneX, \topY-\group*\groupSpacing+\pairOffset) (s1-\group-a) {};
    \node[node] at (\stageOneX, \topY-\group*\groupSpacing-\pairOffset) (s1-\group-b) {};
}

\node[stage] at (\stageTwoX, -1.2) (stageLabel2) {Stage 2};
\node [F] at (\stageTwoX, -1.6) {$F_2=4$};
\foreach \group in {0,1,2,3} {
    \node[node] at (\stageTwoX, \topY-\group*\groupSpacing+\pairOffset) (s2-\group-a) {};
    \node[node] at (\stageTwoX, \topY-\group*\groupSpacing-\pairOffset) (s2-\group-b) {};
}

\node[stage] at (\stageThreeX, -1.2) (stageLabel3) {Stage 3};
\foreach \group in {0,1,2,3} {
    \node[node] at (\stageThreeX, \topY-\group*\groupSpacing+\pairOffset) (s3-\group-a) {};
    \node[node] at (\stageThreeX, \topY-\group*\groupSpacing-\pairOffset) (s3-\group-b) {};
}

\foreach \n/\label in {0/1, 1/2, 2/3, 3/4} {
    \node[font=\small\bfseries] at (\stageThreeX+0.7, \topY-\n*\groupSpacing) {Node \label};
}

\pgfmathsetmacro{\dashOneY}{\topY-0.5*\groupSpacing} 
\pgfmathsetmacro{\dashTwoY}{\topY-1.5*\groupSpacing} 
\pgfmathsetmacro{\dashThreeY}{\topY-2.5*\groupSpacing} 

\pgfmathsetmacro{\dashStartX}{\stageZeroX-0.3}
\pgfmathsetmacro{\dashEndX}{\stageThreeX+0.3}

\draw[dashline] (\dashStartX, \dashOneY) -- (\dashEndX, \dashOneY);
\draw[dashline] (\dashStartX, \dashTwoY) -- (\dashEndX, \dashTwoY);
\draw[dashline] (\dashStartX, \dashThreeY) -- (\dashEndX, \dashThreeY);

\node[blueNode] at (\stageOneX, \topY-1*\groupSpacing+\pairOffset) {};
\node[blueNode] at (\stageOneX, \topY-1*\groupSpacing-\pairOffset) {};

\node[blueNode] at (\stageTwoX, \topY-2*\groupSpacing+\pairOffset) {};
\node[blueNode] at (\stageTwoX, \topY-2*\groupSpacing-\pairOffset) {};

\node[blueNode] at (\stageThreeX, \topY-3*\groupSpacing+\pairOffset) {};
\node[blueNode] at (\stageThreeX, \topY-3*\groupSpacing-\pairOffset) {};

\newcommand{\drawRedX}[2]{
    \draw[redX] (#1-0.2, #2+0.3) -- (#1+0.2, #2-0.3);
    \draw[redX] (#1-0.2, #2-0.3) -- (#1+0.2, #2+0.3);
}

\drawRedX{\stageZeroX}{\topY-1*\groupSpacing}

\drawRedX{\stageOneX}{\topY-2*\groupSpacing}

\drawRedX{\stageTwoX}{\topY-3*\groupSpacing}

%edges
\foreach \i in {0,1,2,3} {
    \foreach \j in {0,1,2,3} {
        \draw[gray!60, line width=0.3pt, opacity=0.3] (m1-\i) -- (s0-\j-a);
        \draw[gray!60, line width=0.3pt, opacity=0.3] (m1-\i) -- (s0-\j-b);
    }
}
\tikzset{
    blackEdge/.style={black!60, line width=0.8pt},
    blueEdge/.style={blue!50, line width=0.8pt},
    edgeLabel/.style={midway, font=\tiny, inner sep=0.5pt, sloped, above=1.5pt}
}

\draw[blackEdge] (s0-0-a) -- (s1-0-a);
\draw[blackEdge] (s0-0-b) -- (s1-0-b);
\draw[blackEdge] (s1-0-a) -- (s2-0-a);
\draw[blackEdge] (s1-0-b) -- (s2-0-b);
\draw[blackEdge] (s2-0-a) -- (s3-0-a);
\draw[blackEdge] (s2-0-b) -- (s3-0-b);

\draw[blackEdge] (s1-1-a) -- (s2-1-a);
\draw[blackEdge] (s1-1-b) -- (s2-1-b);
\draw[blackEdge] (s2-1-a) -- (s3-1-a);
\draw[blackEdge] (s2-1-b) -- (s3-1-b);

\draw[blackEdge] (s0-2-a) -- (s1-2-a);
\draw[blackEdge] (s0-2-b) -- (s1-2-b);
\draw[blackEdge] (s2-2-a) -- (s3-2-a);
\draw[blackEdge] (s2-2-b) -- (s3-2-b);

\draw[blackEdge] (s0-3-a) -- (s1-3-a);
\draw[blackEdge] (s0-3-b) -- (s1-3-b);
\draw[blackEdge] (s1-3-a) -- (s2-3-a);
\draw[blackEdge] (s1-3-b) -- (s2-3-b);

\draw[blueEdge] (s0-0-a) -- node[edgeLabel, pos=0.5, above=1pt] {$b_0(1,1),p_0(1)=1$} (s1-1-a);
\draw[blueEdge] (s0-0-a) -- node[edgeLabel, pos=0.5, below=1pt] {$b_0(1,2),p_0(1)=1$} (s1-1-b);

\draw[blueEdge] (s0-2-a) -- node[edgeLabel, pos=0.4, above=1pt] {$b_0(3,1),p_0(3)=1$} (s1-1-a);
\draw[blueEdge] (s0-2-a) -- node[edgeLabel, pos=0.4, below=1pt] {$b_0(3,2),p_0(3)=1$} (s1-1-b);

\draw[blueEdge] (s0-3-b) -- node[edgeLabel, pos=0.4, above=1pt] {$b_0(4,1),p_0(4)=2$} (s1-1-a);
\draw[blueEdge] (s0-3-b) -- node[edgeLabel, pos=0.4, below=1pt] {$b_0(4,2),p_0(4)=2$} (s1-1-b);

\draw[blueEdge] (s1-0-b) --  (s2-2-a);
\draw[blueEdge] (s1-0-b) --  (s2-2-b);
\draw[blueEdge] (s1-1-a) --  (s2-2-a);
\draw[blueEdge] (s1-1-a) --  (s2-2-b);
\draw[blueEdge] (s1-3-a) --  (s2-2-a);
\draw[blueEdge] (s1-3-a) --  (s2-2-b);

\draw[blueEdge] (s2-0-a) --  (s3-3-a);
\draw[blueEdge] (s2-0-a) --  (s3-3-b);
\draw[blueEdge] (s2-1-b) --  (s3-3-a);
\draw[blueEdge] (s2-1-b) --  (s3-3-b);
\draw[blueEdge] (s2-2-a) --  (s3-3-a);
\draw[blueEdge] (s2-2-a) --  (s3-3-b);
\end{tikzpicture}
	\caption{Example of a signal flow graph for a distributed system containing $n=4$ nodes, each storing $\alpha=2$ packets.}
	\label{fig:graph}
\end{figure}

Figure~\ref{fig:graph_wrong} shows a signal flow graph with the same code parameters, but with a different choice function. In this example, the choice functions of all helper nodes from stages $0$ to $3$ are selected to be $1$. The green and red edges in the figure illustrate an attempt to construct a linking, traced backward from nodes $3$ and $4$ at stage $3$. We observe that at most three vertex-disjoint paths can be found from these vertices to the source vertices at stage $-1$. The red edges indicate that any fourth path starting from the remaining vertex must merge with one of the green paths when traced backward from stage $3$ to stage $-1$. This example shows that the choice of which packet is sent to the newcomer is crucial; the recovery property may fail if the choice function is not selected appropriately.

\begin{figure}[ht]
	\centering
	\begin{tikzpicture}[
    node/.style={draw, circle, fill=black, minimum size=2pt, inner sep=0pt},
    blueNode/.style={draw=blue, circle, fill=blue, minimum size=2pt, inner sep=0pt},
    stage/.style={font=\small\bfseries, align=center},
    F/.style={font=\small, align=center},
    dashline/.style={gray, dashed, line width=0.5pt},
    redX/.style={red, line width=1pt},
    blackEdge/.style={black!60, line width=0.8pt},
    edgeLabel/.style={font=\tiny, inner sep=0.5pt, sloped},
    blueEdge/.style={green!70!black, line width=0.8pt},
    redEdge/.style={red!60,line width=0.8pt}
]

\def\gS{1.6}   % groupSpacing
\def\pO{0.2}   % pairOffset
\def\topY{4.8} % 3*1.6

% x positions
\def\xM{0}
\def\xA{2.5}
\def\xB{6.25}
\def\xC{8.75}
\def\xD{11.25}

% Stage labels
\node[stage] at (\xM,-1.2) {Stage $-1$};
\node[stage] at (\xA,-1.2) {Stage 0};  \node[F] at (\xA,-1.6) {$F_0=2$};
\node[stage] at (\xB,-1.2) {Stage 1};  \node[F] at (\xB,-1.6) {$F_1=3$};
\node[stage] at (\xC,-1.2) {Stage 2};  \node[F] at (\xC,-1.6) {$F_2=4$};
\node[stage] at (\xD,-1.2) {Stage 3};
\foreach \n/\lbl in {0/1,1/2,2/3,3/4}{
    \node[font=\small\bfseries] at (\xD+0.7, \topY-\n*\gS) {Node \lbl};
}

% Stage -1
\foreach \g in {0,1,2,3}{ \node[node] at (\xM,\topY-\g*\gS) (m-\g) {}; }

% Stage 0 (node 2 failed: group 1 gets red X, no newcomer)
\foreach \g in {0,1,2,3}{
    \node[node] at (\xA,\topY-\g*\gS+\pO) (a-\g-u) {};
    \node[node] at (\xA,\topY-\g*\gS-\pO) (a-\g-d) {};
}

% Stage 1 (node 3 failed: group 2 gets red X; group 1 = newcomer)
\foreach \g in {0,2,3}{
    \node[node]     at (\xB,\topY-\g*\gS+\pO) (b-\g-u) {};
    \node[node]     at (\xB,\topY-\g*\gS-\pO) (b-\g-d) {};
}
\node[blueNode] at (\xB,\topY-1*\gS+\pO) (b-1-u) {};
\node[blueNode] at (\xB,\topY-1*\gS-\pO) (b-1-d) {};

% Stage 2 (node 4 failed: group 3 gets red X; group 2 = newcomer)
\foreach \g in {0,1,3}{
    \node[node]     at (\xC,\topY-\g*\gS+\pO) (c-\g-u) {};
    \node[node]     at (\xC,\topY-\g*\gS-\pO) (c-\g-d) {};
}
\node[blueNode] at (\xC,\topY-2*\gS+\pO) (c-2-u) {};
\node[blueNode] at (\xC,\topY-2*\gS-\pO) (c-2-d) {};

% Stage 3 (group 3 = newcomer)
\foreach \g in {0,1,2}{
    \node[node]     at (\xD,\topY-\g*\gS+\pO) (d-\g-u) {};
    \node[node]     at (\xD,\topY-\g*\gS-\pO) (d-\g-d) {};
}
\node[blueNode] at (\xD,\topY-3*\gS+\pO) (d-3-u) {};
\node[blueNode] at (\xD,\topY-3*\gS-\pO) (d-3-d) {};

% Dashed separator lines
\foreach \y in {0.5,1.5,2.5}{
    \draw[dashline] (\xA-0.3,\topY-\y*\gS) -- (\xD+0.3,\topY-\y*\gS);
}

% Red X marks (failed nodes)
\newcommand{\redx}[2]{\draw[redX](#1-0.2,#2+0.3)--(#1+0.2,#2-0.3);\draw[redX](#1-0.2,#2-0.3)--(#1+0.2,#2+0.3);}
\redx{\xA}{\topY-1*\gS}
\redx{\xB}{\topY-2*\gS}
\redx{\xC}{\topY-3*\gS}

% Initialization edges (gray, stage -1 -> stage 0)
\foreach \i in {0,1,2,3}{
    \foreach \g in {0,1,2,3}{
        \draw[gray!60, line width=0.3pt, opacity=0.3] (m-\i) -- (a-\g-u);
        \draw[gray!60, line width=0.3pt, opacity=0.3] (m-\i) -- (a-\g-d);
    }
}

\draw[blueEdge] (m-0) -- (a-0-u); \draw[blueEdge] (m-2) -- (a-2-u); 
\draw[blueEdge] (m-3) -- (a-3-u);
% k-nodes
\node[draw=orange, minimum width=8pt, minimum height=18pt, inner sep=0pt, line width=1pt]
    at (\xD, \topY-2*\gS) {};
\node[draw=orange, minimum width=8pt, minimum height=18pt, inner sep=0pt, line width=1pt]
    at (\xD, \topY-3*\gS) {};

% Carry-forward edges (node 1: group 0, all stages)
\draw[blueEdge] (a-0-u)--(b-0-u); \draw[blackEdge] (a-0-d)--(b-0-d);
\draw[blueEdge] (b-0-u)--(c-0-u); \draw[blackEdge] (b-0-d)--(c-0-d);
\draw[blackEdge] (c-0-u)--(d-0-u); \draw[blackEdge] (c-0-d)--(d-0-d);

% node 2 (group 1): fails at stage 0, alive stage 1->2->3
\draw[blueEdge] (b-1-u)--(c-1-u); \draw[blackEdge] (b-1-d)--(c-1-d);
\draw[blackEdge] (c-1-u)--(d-1-u); \draw[blackEdge] (c-1-d)--(d-1-d);

% node 3 (group 2): alive stage 0->1, fails at stage 1, alive stage 2->3
\draw[blackEdge] (a-2-u)--(b-2-u); \draw[blackEdge] (a-2-d)--(b-2-d);
\draw[blueEdge] (c-2-u)--(d-2-u); \draw[redEdge] (c-2-d)--(d-2-d);

% node 4 (group 3): alive stage 0->1->2, fails at stage 2
\draw[blueEdge] (a-3-u)--(b-3-u); \draw[blackEdge] (a-3-d)--(b-3-d);
\draw[blackEdge] (b-3-u)--(c-3-u); \draw[blackEdge] (b-3-d)--(c-3-d);

% Repair edges, all p_t(i)=1 (helpers send upper sub-node)
% Stage 0->1: repair node 2 (group 1); helpers = nodes 1,3,4 (groups 0,2,3), all send upper
\draw[blackEdge] (a-0-u) -- node[edgeLabel,pos=0.6,above=1pt]{$p_0(1){=}1$} (b-1-u);
\draw[blackEdge] (a-0-u) -- (b-1-d);
\draw[blueEdge] (a-2-u) -- node[edgeLabel,pos=0.6,above=1pt]{$p_0(3){=}1$} (b-1-u);
\draw[blackEdge] (a-2-u) -- (b-1-d);
\draw[blackEdge] (a-3-u) -- node[edgeLabel,pos=0.4,above=1pt]{$p_0(4){=}1$} (b-1-u);
\draw[blackEdge] (a-3-u) -- (b-1-d);

% Stage 1->2: repair node 3 (group 2); helpers = nodes 1,2,4 (groups 0,1,3), all send upper
\draw[blackEdge] (b-0-u) -- (c-2-u); \draw[redEdge] (b-0-u) -- (c-2-d);
\draw[blackEdge] (b-1-u) -- (c-2-u); \draw[redEdge] (b-1-u) -- (c-2-d);
\draw[blueEdge] (b-3-u) -- (c-2-u); \draw[redEdge] (b-3-u) -- (c-2-d);

% Stage 2->3: repair node 4 (group 3); helpers = nodes 1,2,3 (groups 0,1,2), all send upper
\draw[blueEdge] (c-0-u) -- (d-3-u); \draw[blackEdge] (c-0-u) -- (d-3-d);
\draw[blackEdge] (c-1-u) -- (d-3-u); \draw[blueEdge] (c-1-u) -- (d-3-d);
\draw[blackEdge] (c-2-u) -- (d-3-u); \draw[blackEdge] (c-2-u) -- (d-3-d);

\end{tikzpicture}
	\caption{Example of a signal flow graph with inadequate choice functions.}
	\label{fig:graph_wrong}
\end{figure}

Figure~\ref{fig:graph_full} shows a signal flow graph with the same general structure but without the requirements of access-optimality and help-by-transfer. At each stage, all vertices of every helper node are connected to all vertices of the newcomer. Under this configuration, the construction of functional-repair regenerating codes is called ``path-weaving'' in~\cite{Wu10}. The contribution of this paper is to show that path-weaving remains possible in a much sparser signal flow graph, where each helper contributes only one stored packet during repair.

\begin{figure}[ht]
	\centering
	\begin{tikzpicture}[node/.style={draw, circle, fill=black, minimum size=2pt, inner sep=0pt},
    blueNode/.style={draw=blue, circle, fill=blue, minimum size=2pt, inner sep=0pt},
    stage/.style={font=\small\bfseries, align=center},
    F/.style={font=\small, align=center},
    dashline/.style={gray, dashed, line width=0.5pt},
    redX/.style={red, line width=1pt},
    blackEdge/.style={black!70, line width=0.8pt},
    repairEdge/.style={blue!70, line width=0.8pt}
]

\def\gS{1.6}
\def\pO{0.2}
\def\topY{4.8}
\def\xM{0}
\def\xA{2.5}
\def\xB{6.25}
\def\xC{8.75}
\def\xD{11.25}

% Stage labels
\node[stage] at (\xM,-1.2) {Stage $-1$};
\node[stage] at (\xA,-1.2) {Stage 0};  \node[F] at (\xA,-1.6) {$F_0=2$};
\node[stage] at (\xB,-1.2) {Stage 1};  \node[F] at (\xB,-1.6) {$F_1=3$};
\node[stage] at (\xC,-1.2) {Stage 2};  \node[F] at (\xC,-1.6) {$F_2=4$};
\node[stage] at (\xD,-1.2) {Stage 3};
\foreach \n/\lbl in {0/1,1/2,2/3,3/4}{
    \node[font=\small\bfseries] at (\xD+0.7,\topY-\n*\gS) {Node \lbl};
}

% Stage -1
\foreach \g in {0,1,2,3}{ \node[node] at (\xM,\topY-\g*\gS) (m-\g) {}; }

% Stage 0
\foreach \g in {0,1,2,3}{
    \node[node] at (\xA,\topY-\g*\gS+\pO) (a-\g-u) {};
    \node[node] at (\xA,\topY-\g*\gS-\pO) (a-\g-d) {};
}

% Stage 1 (group 1 = newcomer)
\foreach \g in {0,2,3}{
    \node[node]     at (\xB,\topY-\g*\gS+\pO) (b-\g-u) {};
    \node[node]     at (\xB,\topY-\g*\gS-\pO) (b-\g-d) {};
}
\node[blueNode] at (\xB,\topY-1*\gS+\pO) (b-1-u) {};
\node[blueNode] at (\xB,\topY-1*\gS-\pO) (b-1-d) {};

% Stage 2 (group 2 = newcomer)
\foreach \g in {0,1,3}{
    \node[node]     at (\xC,\topY-\g*\gS+\pO) (c-\g-u) {};
    \node[node]     at (\xC,\topY-\g*\gS-\pO) (c-\g-d) {};
}
\node[blueNode] at (\xC,\topY-2*\gS+\pO) (c-2-u) {};
\node[blueNode] at (\xC,\topY-2*\gS-\pO) (c-2-d) {};

% Stage 3 (group 3 = newcomer)
\foreach \g in {0,1,2}{
    \node[node]     at (\xD,\topY-\g*\gS+\pO) (d-\g-u) {};
    \node[node]     at (\xD,\topY-\g*\gS-\pO) (d-\g-d) {};
}
\node[blueNode] at (\xD,\topY-3*\gS+\pO) (d-3-u) {};
\node[blueNode] at (\xD,\topY-3*\gS-\pO) (d-3-d) {};

% Dashed separators
\foreach \y in {0.5,1.5,2.5}{
    \draw[dashline] (\xA-0.3,\topY-\y*\gS) -- (\xD+0.3,\topY-\y*\gS);
}

% Red X
\newcommand{\redx}[2]{
    \draw[redX](#1-0.2,#2+0.3)--(#1+0.2,#2-0.3);
    \draw[redX](#1-0.2,#2-0.3)--(#1+0.2,#2+0.3);
}
\redx{\xA}{\topY-1*\gS}
\redx{\xB}{\topY-2*\gS}
\redx{\xC}{\topY-3*\gS}

% Initialization edges (gray)
\foreach \i in {0,1,2,3}{
    \foreach \g in {0,1,2,3}{
        \draw[gray!60,line width=0.3pt,opacity=0.3] (m-\i)--(a-\g-u);
        \draw[gray!60,line width=0.3pt,opacity=0.3] (m-\i)--(a-\g-d);
    }
}

% Carry-forward edges
\draw[blackEdge](a-0-u)--(b-0-u); \draw[blackEdge](a-0-d)--(b-0-d);
\draw[blackEdge](b-0-u)--(c-0-u); \draw[blackEdge](b-0-d)--(c-0-d);
\draw[blackEdge](c-0-u)--(d-0-u); \draw[blackEdge](c-0-d)--(d-0-d);

\draw[blackEdge](b-1-u)--(c-1-u); \draw[blackEdge](b-1-d)--(c-1-d);
\draw[blackEdge](c-1-u)--(d-1-u); \draw[blackEdge](c-1-d)--(d-1-d);

\draw[blackEdge](a-2-u)--(b-2-u); \draw[blackEdge](a-2-d)--(b-2-d);
\draw[blackEdge](c-2-u)--(d-2-u); \draw[blackEdge](c-2-d)--(d-2-d);

\draw[blackEdge](a-3-u)--(b-3-u); \draw[blackEdge](a-3-d)--(b-3-d);
\draw[blackEdge](b-3-u)--(c-3-u); \draw[blackEdge](b-3-d)--(c-3-d);

% Repair edges (blue): every helper sub-node -> every newcomer sub-node
% Stage 0->1: helpers = groups 0,2,3; newcomer = group 1
\foreach \h in {0,2,3}{
    \foreach \hs in {u,d}{
        \draw[repairEdge](a-\h-\hs)--(b-1-u);
        \draw[repairEdge](a-\h-\hs)--(b-1-d);
    }
}

% Stage 1->2: helpers = groups 0,1,3; newcomer = group 2
\foreach \h in {0,1,3}{
    \foreach \hs in {u,d}{
        \draw[repairEdge](b-\h-\hs)--(c-2-u);
        \draw[repairEdge](b-\h-\hs)--(c-2-d);
    }
}

% Stage 2->3: helpers = groups 0,1,2; newcomer = group 3
\foreach \h in {0,1,2}{
    \foreach \hs in {u,d}{
        \draw[repairEdge](c-\h-\hs)--(d-3-u);
        \draw[repairEdge](c-\h-\hs)--(d-3-d);
    }
}

\end{tikzpicture}
	\caption{ Signal flow graph without access-optimality}
	\label{fig:graph_full}
\end{figure}

\begin{remark}
The construction of the signal flow graph depends on the order in which the storage nodes fail. That is, it depends on the failure sequence $(F_t)_{t\geq 0}$. For a different failure sequence, the structure of the signal flow graph may be different.
\end{remark}

\begin{remark}
Recall that the above discussion assumes $\beta=1$. For $\beta>1$, provided that the parameters are scaled accordingly, a signal flow graph with parameters $(n,k,\alpha',\beta,B')$ can be obtained by taking $\beta$ parallel copies of a signal flow graph with parameters
\[
    \left(n,k,\frac{\alpha'}{\beta},1,\frac{B'}{\beta}\right),
\]
and grouping the corresponding copies of each storage node together. In the resulting graph, each helper transmits one packet in each copy, and hence transmits a total of $\beta$ packets during repair. It is straightforward to verify that the resulting graph preserves the corresponding linking and recovery properties of the underlying $\beta=1$ construction. Therefore, it is sufficient to discuss signal flow graphs with $\beta=1$.
\end{remark}
 
Although the signal flow graph may be infinite, the analysis of the $N$-th node failure and repair only involves the vertices and edges in stages $-1,0,\ldots,N$. It therefore suffices to consider the finite subgraph induced by these stages. We call this finite graph the {\em signal flow graph truncated to stage $N$}, or simply the truncated graph when $N$ is clear from the context.

The truncation is used only to isolate the finite structure relevant to the analysis of a given stage. The construction itself is not tied to any fixed truncated graph, and extends stage by stage without any prescribed upper bound on the number of repairs.

Within each truncated graph, we first determine the choice functions while leaving the local encoding coefficients unspecified. This is because the relevant combinatorial structure depends only on the existence of linkings in the directed graph, and not on the labels assigned to the edges. Once the combinatorial structure has been determined, we use the fact that the signal flow graph induces a gammoid. Since finite gammoids are linearly representable over sufficiently large finite fields, the required combinatorial independences can then be realized as linear independences of global encoding vectors. This realizes the signal flow graph as a discrete-time linear dynamical system.

We use $G=(\sV,\sE)$ to denote the {\em unlabeled} signal flow graph. Since the vertices at stage $-1$ represent the information source, it is convenient to reverse the directions of all edges and consider the opposite graph. We denote by $G^{\mathrm{op}}$ the graph obtained from $G$ by reversing the direction of every edge. We then define the relevant gammoids on this opposite graph.

We use $(i,j)_t$ to denote the stage-indexed label corresponding to the vertex $v_t(i,j)$, where $(i,j)\in\sJ$ and $t\geq 0$. The subscript~$t$ indicates that the index pertains to stage $t$. For a set $\mathcal{I}$ of such stage-indexed labels, let $\mathcal{V}(\mathcal{I})$ denote the corresponding set of vertices:
\[
    \mathcal{V}(\mathcal{I})
    :=
    \{v_t(i,j): (i,j)_t\in\mathcal{I}\}.
\]
Let $\mathcal{G}(\mathcal{I})$ denote the family of global encoding vectors corresponding to $\mathcal{I}$, where vectors carrying distinct labels are retained as distinct members even if they are equal. 

For each stage $t\geq 0$, define the index set
\[
    \mathcal{J}_t
    :=
    \{(i,j)_t:\, i=1,2,\ldots,n,\ j=1,2,\ldots,\alpha\}.
\]
The corresponding vertices and global encoding vectors at stage $t$ are then denoted respectively by
\[
    \mathcal{V}_t := \mathcal{V}(\mathcal{J}_t),
    \qquad
    \mathcal{G}_t := \mathcal{G}(\mathcal{J}_t).
\]

The vertices at stage $-1$ are elements of the set $\mathcal{V}_{-1}$.

For a set $\mathcal{I}$ of stage-indexed labels, define its restriction to stage $t$ by $$
\mathcal{I}_t:=\mathcal{I}\cap \mathcal{J}_t.$$

\begin{definition}
For each $t\geq 0$, consider the strict gammoid
$L(G^{\mathrm{op}},\mathcal{V}_{-1})$,
where the target set is the set of source vertices $\mathcal{V}_{-1}$. We define $\mathscr{M}_t$ to be the restriction of this strict gammoid to the subset
$\mathcal{V}_t\cup \mathcal{V}_{t+1}$.
That is,
\[
    \mathscr{M}_t  :=
    \left.
    L(G^{\mathrm{op}},\mathcal{V}_{-1})
    \right|_{\mathcal{V}_t\cup \mathcal{V}_{t+1}}.
\]
We denote by $\mathscr{C}_t$ the collection of independent sets of $\mathscr{M}_t$.
\end{definition}

Thus, a subset $\mathcal{S}\subseteq \mathcal{V}_t\cup \mathcal{V}_{t+1}$ is independent in $\mathscr{M}_t$ if and only if $\mathcal{S}$, regarded as a set of vertices in $G^{\mathrm{op}}$, can be linked into $\mathcal{V}_{-1}$.

Throughout the remainder of the paper, edges are described in the original signal flow graph $G$, whereas linkings are considered in the opposite graph $G^{\mathrm{op}}$.

Since the edges of $G^{\mathrm{op}}$ connect only consecutive stages, a linking can be decomposed stage by stage. In particular, a subset of vertices in $\mathcal{V}_{t+1}$ can be linked into $\mathcal{V}_{-1}$ if and only if it can first be linked onto a subset of $\mathcal{V}_t$, which can then be linked into $\mathcal{V}_{-1}$. Based on this observation, the independence of any set
\[
    \mathcal{U}\subseteq \mathcal{V}_t\cup\mathcal{V}_{t+1}
\]
in the gammoid $\mathscr{M}_t$ can be reduced to the analysis of a vertex set at stage $t$.

The following theorem formalizes this property.

\begin{theorem}
\label{thm: graph_property}
Let $\mathcal{U}\subseteq \mathcal{V}_t\cup\mathcal{V}_{t+1}$ be a subset of the ground set of $\mathscr{M}_t$. The following statements are equivalent.

\begin{enumerate}
\item $\mathcal{U}$ is independent in $\mathscr{M}_t$.

\item There exists a set $\mathcal{W}\subseteq \mathcal{V}_t\setminus \mathcal{U}$ such that the following two conditions hold:
\begin{enumerate}
\item there is a perfect matching from $\mathcal{W}$ onto $\mathcal{U}\cap \mathcal{V}_{t+1}$, using the corresponding edges of the graph~$G$;

\item the set $(\mathcal{U}\cap \mathcal{V}_t)\cup \mathcal{W}$ can be linked into $\mathcal{V}_{-1}$ in $G^{\mathrm{op}}$.
\end{enumerate}
\end{enumerate}
\end{theorem}

\begin{proof}
    See Appendix \ref{apd: 3}.
\end{proof}

\section{From Regenerating Codes to Linking Conditions}
\label{sec: reduction}

This section reduces the code construction to a signal flow graph satisfying certain linking conditions and identifies the requirements under which the construction supports an unbounded number of repairs. 

Recall that the construction aims to maintain the $(n,k)$ recovery property after each repair, while ensuring that every repair is access-optimal and help-by-transfer. These requirements correspond to linear independence conditions on the global encoding vectors. 

To reduce the linear independence to linking conditions, we first consider a natural bijection
\[
    \theta: v_m(i,j)\mapsto (i,j)_m
\]
from the ground set $\mathcal{V}_t\cup\mathcal{V}_{t+1}$ of $\mathscr{M}_t$ to the index set
\[
    \mathcal{J}_t\cup\mathcal{J}_{t+1},
\]
which is the ground set of the vector matroid formed by the global encoding vectors
$\mathcal{G}_t\cup\mathcal{G}_{t+1}$. Therefore, if the gammoid $\mathscr{M}_t$ and this vector matroid are isomorphic under $\theta$, then combinatorial independence in the gammoid corresponds precisely to linear independence of the associated global encoding vectors. 

Our construction is built on this isomorphism. We emphasize that the two types of components in the signal flow graph determine the two kinds of independence respectively. The choice functions determine the edge structure of the graph and thereby establish the combinatorial independence in the gammoid, whereas the local encoding coefficients, serving as the edge labels, determine whether the induced vector matroid provides a linear representation of the gammoid.

We first translate the recovery and repair requirements into combinatorial linking conditions on the signal flow graph. These are the following two linking conditions.

\begin{definition}[Recovery linking condition]
A signal flow graph is said to satisfy the {\em recovery linking condition} at stage $t+1$ if, for every set $\mathbf{K}$ of $k$ distinct storage nodes, the vertex set at stage $t+1$ corresponding to the nodes in $\mathbf{K}$ contains an independent subset of size $B$ in the gammoid~$\mathscr{M}_t$.
\end{definition}

\begin{definition}[Repair linking condition]
A signal flow graph is said to satisfy the {\em repair linking condition} at stage $t+1$ if the following two conditions hold:
\begin{enumerate}
\item {\em Positive part:} the set
\[
    \{v_t(i,p_t(i)):\, i\neq F_t\}
\]
can be linked into $\mathcal{V}_{-1}$.

\item {\em Negative part:} for every $j\in[\alpha]$, the set
\[
    \{v_t(i,p_t(i)):\, i\neq F_t\}\cup \{v_{t+1}(F_t,j)\}
\]
cannot be linked into $\mathcal{V}_{-1}$.
\end{enumerate}
\end{definition}

The positive part of the repair linking condition says that the set
$\{v_t(i,p_t(i)):\, i\neq F_t\}$ is independent in the gammoid $\mathscr{M}_t$, while the negative part says that $v_{t+1}(F_t,j)$ is dependent on this set. Recall from Lemma~\ref{lemma:B_larger_than_n_minus_1} that $B\geq n-1$. Hence, the size of the set $\{v_t(i,p_t(i)):\, i\neq F_t\}$ is at most the size of $\mathcal{V}_{-1}$.

To establish this translation, we develop a framework for incremental linear representations of the gammoids induced by the signal flow graph. The basic prerequisite in our construction is the linear representability of the relevant gammoids. As discussed in Section~\ref{sec: matroid}, every finite gammoid is linearly representable over a sufficiently large finite field. This result, however, is existential. It assumes that the gammoid is specified in advance, and then assigns vectors to its vertices so that combinatorial independence is represented by linear independence. In our setting, the underlying graph grows over time, giving rise to a sequence of gammoids
\[
    \mathscr{M}_0,\mathscr{M}_1, \mathscr{M}_2, \ldots
\]
that appear successively. Consequently, the linear representation cannot be constructed once and for all; rather, it must be built incrementally.

In view of this, we establish a theorem that characterizes a single-step linear representation of a gammoid under the constraints imposed by the discrete-time dynamical system. The theorem also provides an explicit requirement on the field size.

\begin{theorem}
\label{thm: linear_repre}
Suppose that the size $q$ of the finite field $\mathbb{F}_q$ satisfies
\[
	q> \binom{n\alpha}{B}-\binom{(n-1)\alpha}{B}.
\]

For any $t\geq 1$, if the vector matroid formed by $\mathcal{G}_{t-1}\cup\mathcal{G}_t$ is isomorphic to $\mathscr{M}_{t-1}$ under the natural bijection $\theta$ over $\mathbb{F}_q$, then the local encoding coefficients at stage $t$ can be chosen so that the vector matroid formed by $\mathcal{G}_t\cup\mathcal{G}_{t+1}$ is isomorphic to $\mathscr{M}_t$ under the bijection $\theta$.
\end{theorem}

\begin{proof}
    See Section \ref{sec: linear}.
\end{proof}

We note that the bound on the finite-field size in Theorem~\ref{thm: linear_repre} is independent of the stage index $t$. Moreover, $\mathcal{G}_{t+1}$ depends only on $\mathcal{G}_t$ and the corresponding graph structure between stages $t$ and $t+1$. This allows us to apply the theorem recursively, thereby constructing linear representations of the unbounded sequence of gammoids incrementally. The following corollary summarizes this process.

\begin{coro}
\label{coro: linear_repre}
Suppose that the finite field size satisfies the condition in Theorem~\ref{thm: linear_repre}. Then there exists a sequence
$\mathcal{G}_0,\mathcal{G}_1,\ldots$ of indexed families of global encoding vectors
over the fixed finite field $\FF{q}$ such that, for each $t\geq 0$, the gammoid $\mathscr{M}_t$ is isomorphic to the vector matroid formed by $\mathcal{G}_t\cup\mathcal{G}_{t+1}$ under the natural bijection between vertices and global encoding vectors.
\end{coro}

We will prove this corollary as a consequence of Theorem~\ref{thm: linear_repre} in Section~\ref{sec: linear}.

With the isomorphism established in Theorem \ref{thm: linear_repre} and Corollary \ref{coro: linear_repre}, we formalize the translation of the recovery and repair properties of the regenerating code into the two linking conditions defined at the beginning of this section.

\begin{theorem}
\label{thm:linking_conditions_to_code}
The following statements hold.
\begin{enumerate}
\item If a signal flow graph satisfies the recovery linking condition at every stage, then the sequence of global encoding vector families obtained from Corollary~\ref{coro: linear_repre} forms a linear regenerating code satisfying the $(n,k)$ recovery property at every stage.

\item If a signal flow graph satisfies the repair linking condition at every stage, then the sequence of global encoding vector families obtained from Corollary~\ref{coro: linear_repre} forms an access-optimal help-by-transfer linear regenerating code.

\end{enumerate}
\end{theorem}

\begin{proof} (i)
By the initialization at stage $0$, the $(n,k)$ recovery property at stage $0$ is satisfied.

Fix any $t\geq 0$. The recovery linking condition at stage $t+1$ guarantees that, for any set $\mathbf{K}$ of $k$ storage nodes, the corresponding vertex set at stage $t+1$ contains an independent subset of size $B$ in $\mathscr{M}_t$. By the isomorphism in Corollary~\ref{coro: linear_repre}, the corresponding global encoding vectors are linearly independent. Since the ambient vector space is $\FF{q}^B$, these $B$ linearly independent vectors form a basis of $\FF{q}^B$. Therefore, the original data file can be recovered from the coded packets stored in the nodes in $\mathbf{K}$. This proves the $(n,k)$ recovery property at stage $t+1$.    

%Fix any $t\geq 0$, the recovery linking condition at stage $t+1$ guarantees that the vertex set, corresponding to the nodes in $\mathbf{K}$ at stage $t+1$, contains an independent subset of size $B$. The corresponding global encoding vectors in the vector matroid is linearly independent by the isomorphism, and hence form a basis of $\FF{q}^B$. We can therefore recover the raw data from the associated coded packets.
    
%    By Theorem \ref{thm: linear_repre} and Corollary \ref{coro: linear_repre}, there exist local encoding coefficients such that the $B$ global encoding vectors corresponding to this independent subset are linearly independent, therefore $\{\mathbf{e}_{t+1}(i,j)\mid i\in \{i_1,\dots,i_k\}, j\in[\alpha]\}$ can span the whole vector space $\mathbb{F}_q^B$, which is the $(n,k)$ recovery property at stage $t+1$.

(ii) 
For any $t\geq 0$, consider the repair of node $F_t$. Each helper node $i\neq F_t$ reads and transmits the packet
\[
    S_t(i,p_t(i)),
\]
whose global encoding vector is $\be_t(i,p_t(i))$. By the positive part of the repair linking condition, the vertex set
\[
    \{v_t(i,p_t(i)):\, i\neq F_t\}
\]
is independent in $\mathscr{M}_t$. Hence, by the isomorphism in Corollary~\ref{coro: linear_repre}, the corresponding global encoding vectors are linearly independent.

The negative part of the repair linking condition implies that, for each $j=1,2,\ldots,\alpha$, the set
\[
    \{v_t(i,p_t(i)):\, i\neq F_t\}\cup \{v_{t+1}(F_t,j)\}
\]
is dependent in $\mathscr{M}_t$. Since the first set is independent, adding $v_{t+1}(F_t,j)$ does not increase the rank. Therefore, in the corresponding vector matroid, the global encoding vector associated with $v_{t+1}(F_t,j)$ lies in the span of the global encoding vectors associated with the repair packets
\[
    \{S_t(i,p_t(i)):\, i\neq F_t\}.
\]
Thus, each packet stored in the newcomer can be computed from the packets transmitted by the helpers.

Moreover, each surviving node reads and transmits exactly one stored packet and performs no local computation. Therefore, the repair is help-by-transfer and access-optimal.
\end{proof}

To ensure that the construction supports an unbounded number of repairs, two further requirements should be satisfied for each repair.
\renewcommand{\labelenumi}{\arabic{enumi}.}
\begin{enumerate}
\item \textbf{Fixed finite field}: the entire repair process should be carried out over a single fixed finite field whose size is independent of the stage index $t$;
\item \textbf{Causal finite memory}: each repair depends only on a finite window of the failure history with length bounded independently of $t$. This requirement has two sides: the local encoding coefficient side and the choice function side.
\end{enumerate}

These two requirements are established at different points of the construction. Requirement 1, together with the coefficient side of Requirement 2, is ensured by Theorem~\ref{thm: linear_repre} and Corollary~\ref{coro: linear_repre}. The choice function side of Requirement 2 is deferred to the next section, where we give an algorithm that selects the choice functions incrementally, depending only on a finite history.

By the conclusions above, the code construction reduces to the construction of an unlabeled signal flow graph that satisfies both the recovery linking condition and the repair linking condition at every stage, and whose choice functions are chosen incrementally from a finite history. We establish this construction in the next section.

\section{Admissible Signal Flow Graph}

\label{sec:packet_selection}

This section shows that the linking conditions defined in the previous section can be achieved combinatorially by an {\em admissible signal flow graph}. For any given failure sequence, we provide a stage-by-stage algorithm that selects the choice functions and ensures that the generated signal flow graph satisfies the linking conditions. The algorithm depends on a window of finite history, rather than the entire node failure sequence. In the next section, we will see how to realize an admissible signal flow graph using linear algebra and the vector matroid representation.

%In Section \ref{subsec: recovery} and Section \ref{subsec:repair}, we prove that an admissible signal flow graph satisfies the recovery and repair linking conditions, respectively. We then give the explicit algorithm for selecting choice functions and prove that the generated signal flow graph is admissible. Since the algorithm depends only on at most $\alpha$ distinct nodes rather than the entire failure sequence, the construction supports an unbounded number of repairs as described in the previous section.

We first set up notation for finite windows of the failure sequence and the choice functions on them.

\begin{definition}
For $a\leq b$, we use $\sF{a,b}:=\{F_a,F_{a+1},\dots,F_b\}$ to denote the set of failed nodes corresponding to stages $a,a+1,\dots,b$. Likewise, for each node $i$, we define the set of values of $p_m(i)$ from stage $a$ to $b$ by
$$\sP{a,b}(i):=\{p_m(i)\mid a\leq m\leq b \text{ and } F_m\neq i\}.$$

We adopt the convention that both $\sF{a,b}$ and $
\sP{a,b}(i)$ are empty sets whenever $a>b$.
\end{definition}

The set $\sP{a,b}(i)$ consists of the indices of the packets that are read by node $i$ at the stages from $a$ to $b$. The condition $F_m\neq i$ is included because $p_m(i)$ is undefined when node $i$ fails at stage $m$.

With this notation in place, we now state the central condition imposed on the choice functions.

\begin{definition}[Admissible signal flow graph]
    A signal flow graph is said to be {\em admissible} if, for any two stages $a\leq b$ such that $|\sF{a,b}|\leq \alpha$ and every node $i\in [n] \setminus \sF{a,b}$, we have
\begin{equation} p_a(i)=p_b(i) \iff  F_a=F_b. \label{eq:admissible}
    \end{equation}
\end{definition}

\begin{example}
The signal flow graph in Fig.~\ref{fig:graph} is admissible. We can check that, for $a=0$ and $b=1$, we have $|\sF{0,1}|=2$, and nodes 1 and 4 are not in $\sF{0,1}$. We check that \eqref{eq:admissible} holds by observing $p_0(1)=1 \neq p_1(1)=2$ and $p_0(4) =2 \neq p_1(4) = 1$. Likewise, for $a=1$ and $b=2$, we have $|\sF{1,2}|=2$, and nodes 1 and 2 are not in $\sF{1,2}$. We check that $p_1(1) \neq p_2(1)$ and $p_1(2) \neq p_2(2)$. 

However, the signal flow graph in Fig.~\ref{fig:graph_wrong} is not admissible. 
\end{example}

\begin{remark} We note that the condition of admissibility is $|\sF{a,b}|\leq \alpha$, while the difference $b-a$ may be larger than~$\alpha$. For instance, we may have $1 = F_0=F_1=F_2=\cdots$, that is, node 1 fails at every stage. Then, admissibility requires that $p_a(i)=p_b(i)$ for all nodes $i\neq 1$.
\end{remark}

Both linking conditions in this section rely on admissibility through the following counting consequence, which we will use repeatedly.

\begin{coro}
\label{coro: distinct_choice}
Suppose a signal flow graph is admissible. For any $a\leq b$ such that $|\sF{a,b}|\le\alpha$, and any node 
$i$ such that $i\notin\sF{a,b}$, we have $|\sP{a,b}(i)|=|\sF{a,b}|$.
\end{coro}

\begin{proof}
Let $i$ be a node index that is not in $\sF{a,b}$.
For stages $m,m'$ with $a\le m,m'\le b$, the condition of admissibility implies that $p_m(i)=p_{m'}(i)$ if and only if $F_m=F_{m'}$. Therefore, the map $m\mapsto p_m(i)$ induces a bijection between the distinct values in $\sF{a,b}$ and the distinct values in $\sP{a,b}(i)$. This proves that the number of distinct values in $\sP{a,b}(i)$ is the same as the number of distinct values in $\sF{a,b}$, that is, $|\sP{a,b}(i)|=|\sF{a,b}|$. 
\end{proof}

Admissibility depends on windows containing at most $\alpha$ distinct failed nodes, and a graph satisfies it when every such window does. Truncation at stage $N$ preserves every window ending at or before stage $N$, and admissibility is therefore unaffected by truncation. The conclusion at the truncated stage therefore extends to every stage of the infinite graph.

Throughout, we fix an arbitrary $N$ and an arbitrary failure sequence $(F_0,\dots,F_{N-1})$, and consider the admissible signal flow graph with parameters $(n,k,\alpha,\beta,B)$ truncated at stage $N$. We prove that an admissible signal flow graph truncated at stage $N$ satisfies both linking conditions at stage $N$.

Next, we introduce the notion of $F$-pair. Since the number of nodes is finite and time is unbounded, a node may fail infinitely many times. If we write down the sequence of stage indices when a given node fails, each pair of adjacent numbers in this sequence is called an $F$-pair.

\begin{definition}[$F$-pair]
    Given the sequence of failed nodes $(F_0,F_1,\dots)$, a pair of stage indices $(s,t)$ is called an {\em adjacent identical $F$-pair}, in short {\em $F$-pair}, if 
\begin{enumerate}
\item $0\leq s < t$,
\item  $F_s=F_t$, and  
\item $F_m\neq F_s$ for all $s<m<t$.
\end{enumerate}    
Any such stage $m$ is called an {\em intermediate stage} of the $F$-pair $(s,t)$. 
\end{definition}

\begin{definition}[$F$-pair for truncated signal flow graph]
For a signal flow graph truncated at stage $N$, a pair $(s,N)$ of stage indices is called an {\em $F$-pair} if $F_m\neq F_s$ for all $s<m<N$.
\end{definition}

\begin{example}
In Fig.~\ref{fig:graph},  the signal flow graph is truncated at stage 3. We say that $(0,3)$ is an $F$-pair because $F_m\neq F_0$ for $m=1,2$.
\end{example}

\begin{remark}
    The arguments in Sections \ref{subsec: recovery} and \ref{subsec:repair} will be expressed in terms of $F$-pairs, but from {\em opposite} directions: Section~\ref{subsec: recovery} takes the earlier stage $s$ of an $F$-pair $(s,t)$ as its reference, whereas Section~\ref{subsec:repair} takes the later stage $t$. We nonetheless write every $F$-pair as $(s,t)$ with $s<t$ throughout.
\end{remark}

\subsection{Admissibility implies the recovery linking condition} \label{subsec: recovery}

In this subsection, we prove that an admissible signal flow graph satisfies the recovery linking condition at stage $N$. By Theorem \ref{thm: graph_property}, the existence of a linking of any vertex set at stage $s$ into $\sV_{-1}$ reduces to that of an associated vertex set at stage $s-1$. Applying this reduction inductively from stage $N$ to stage $0$, we reduce the proof to tracking the decreases in the cardinality of the retained set at each stage.

We let $\sK$ denote a set of $k$ distinct node indices.
Recall that the recovery linking condition requires that, starting from the $k\alpha$ vertices at stage $N$ corresponding to $\sK$, a subset of size $B$ can be linked onto $\sV_{-1}$.
To track such a linking, we introduce a vertex subset $\mV{s}\subset \mathcal{V}_s$ for each stage $s\geq 0$.

\begin{definition}[Active vertices] Given a set $\sK$ of $k$ distinct node indices and a nonnegative integer $N$,
we define 
$$\mV{N} :=\{v_N(i,j)\mid i\in\mathbf{K},j\in[\alpha]\}$$ 
and construct $\mV{s}$, for $s<N$, backwards stage by stage: Given $\mV{s+1}$, we let $\mV{s}$ be a maximum-cardinality subset of vertices in stage $s$ subject to the existence of a subset of $\mV{s+1}$ that can be linked onto $\mV{s}$.
\end{definition}

Strictly speaking, the set $\mV{N}$ depends on $\sK$, but we will suppress this dependence for notational simplicity. We note that the active vertices $\mV{s}$ may not be uniquely determined, because there may be more than one choice of subsets that have a maximum-cardinality matching with the vertices in $\mV{s+1}$.

The set $\mV{N}$ contains a subset that can be linked onto $\mV{0}$, with $|\mV{0}|$ maximized. Because each vertex in $\mV{s}$ is matched to a unique vertex in $\mV{s+1}$, we have $|\mV{s}|\leq |\mV{s+1}|$. By the initialization of stage $0$, any vertex subset at stage $0$ of size $B$ can be linked onto $\mathcal{V}_{-1}$. Therefore, if $|\mV{0}|\geq B$ holds for every choice of $\mathbf{K}$, a subset of $\mV{N}$ of size $B$ can be linked onto $\mathcal{V}_{-1}$, and the recovery linking condition follows.

To capture the loss in $|\mV{s}|$ and its compensation, we introduce two quantities relative to $\mV{s+1}$.

\begin{definition}[Failed vertices]
Given $\mV{s+1}$, the {\em failed vertices} at stage $s$ are defined as
$$
\mathcal{D}_s:= \mV{s+1}\cap \{v_{s+1}(F_s,j)\mid j\in[\alpha]\}.
$$
\end{definition}

The vertices in $\mathcal{D}_s$ correspond to packets stored in the newcomer at stage $s+1$.

\begin{definition}[Available nodes] We say that a node $i\neq F_s$ is {\em available} at stage $s$ if $$v_{s+1}(i,p_s(i))\notin \mV{s+1}.$$
We denote 
the set of all available nodes at stage $s$ by $\mathbf{N}_s$.
\end{definition}

In other words, a node $i\neq F_s$ is an available node at stage $s$ if the vertex corresponding to the packet that is transmitted from node $i$ to the new node at stage $s$ is not an active vertex in $\mV{s+1}$.

Here, $|\mathcal{D}_s|$ measures the potential decrease in $|\mV{s}|$, and $|\mathbf{N}_s|$ is the number of nodes that can offset this loss. Note that $\mathcal{D}_s\subset \sV_{s+1}$ is a set of vertices at stage $s+1$, while $\mathbf{N}_s\subset[n]$ is a set of nodes.

We now give a particular backward construction of active vertices. 

\begin{construction}
Given $\mV{s+1}$, the set $\mV{s}$ consists of two parts. The first part is $$\{v_s(i,j)\mid v_{s+1}(i,j)\in\mV{s+1}\setminus\mathcal{D}_s\},
$$
and the second part is an arbitrary subset of $\{v_s(i,p_s(i))\mid i\in\mathbf{N}_s\}$ of size $\min(|\mathcal{D}_s|,|\mathbf{N}_s|)$.  
\end{construction}

In the transition from stage $s$ to $s+1$, only the packets in the failed node $F_s$ are changed.
The first part of $\mV{s}$ corresponds to the packets stored in the surviving nodes. The second part consists of the accessed packets from some available nodes.

The following lemma shows that this construction attains the maximum achievable $|\mV{s}|$ for any given $\mV{s+1}$. Although the set $\mV{s}$ is not unique, the results below hold for any $\mV{s}$ produced by this construction.

\begin{lemma}
   \label{lem: Act_bound}
    Given $\mV{s+1}$, the maximum of $|\mV{s}|$ is $|\mV{s+1}|-\max(0,|\mathcal{D}_s|-|\mathbf{N}_s|)$, which can be achieved by the backward construction.
\end{lemma}

\begin{proof}

    By the construction of the signal flow graph, every vertex $v_{s+1}(i,j)\in\mV{s+1}$ with $i \neq F_s$ has a unique incoming edge from $v_s(i,j)$. If $\mV{s}$ does not contain $v_s(i,j)$, then $\mV{s} \cup \{v_s(i,j)\}$ forms a valid set of active vertices at stage $s$ of strictly larger size. We therefore assume $$\{v_s(i,j)\mid v_{s+1}(i,j)\in\mV{s+1},i\neq F_s\}\subset \mV{s}.$$
    This subset has size $|\mV{s+1}|-|\mathcal{D}_s|$.

    The remaining vertices that $\mV{s}$ may contain are those of the form $v_s(i,p_s(i))$ with $i\neq F_s$, since these are the only vertices in stage $s$ with edges to some $v_{s+1}(F_s,j)$. Such a vertex falls outside the above subset if and only if $i\in\mathbf{N}_s$. Therefore, $|\mV{s}|\leq |\mV{s+1}|-|\mathcal{D}_s|+|\mathbf{N}_s|$. Since $|\mV{s}|\leq |\mV{s+1}|$ by the definition of active vertices, taking the minimum of the two bounds yields the upper bound in the statement. 
     
    Clearly, the backward construction attains this bound.
\end{proof}

Lemma \ref{lem: Act_bound} reduces the proof to bounding $|\mathcal{D}_s|$ and $|\mathbf{N}_s|$, both of which depend on the choice of $\mV{s+1}$. We show that both quantities can be bounded by the structure of the signal flow graph. Since both bounds rely on the same forward-tracing argument, we establish the following lemma first.

\begin{lemma}
    \label{lem: forward_argu}
    For any two stages $a,b$, if a node $i$ satisfies $i\notin\sF{a,b-1}$, $v_b(i,j)\notin\mV{b}$ for some $j\in[\alpha]$, then $v_a(i,j)\in\mV{a}$ only if $j=p_m(i)$ for some $a\leq m<b$.
\end{lemma}
\begin{proof}
    Let $v_a(i,j)\in\mV{a}$. Since $i\notin\sF{a,b-1}$, and $v_b(i,j)\notin\mV{b}$, by the construction of active vertices, there must exist a stage $a\leq m<b$ at which $v_m(i,j)\in\mV{m}$ while $v_{m+1}(i,j)\notin\mV{m+1}$. This implies $v_m(i,j)$ is contained in the second part of $\mV{m}$, and therefore $j=p_m(i)$.
\end{proof}

We first give an upper bound on $\mathcal{D}_s$.

\begin{lemma}[Bound on the size of $\mathcal{D}_s$] 
\label{lem: FV}
    Let $(s,t)$ be an $F$-pair with  $t\neq N$. Then 
    \begin{equation} \mathcal{D}_s\subseteq \{v_{s+1}(F_s,j)\mid  j\in\sP{s+1,t-1}(F_s)\}.
    \label{eq:Ds_bound}
    \end{equation}
\end{lemma}

\begin{proof}
    Since $(s,t)$ is an $F$-pair, $F_s=F_t$ is the failed node at stage $t$, and the construction of active vertices implies that $v_t(F_s,j)\notin\mV{t}$ for every $j\in[\alpha]$. Since $F_s\notin\sF{s+1,t-1}$, for any $v_{s+1}(F_s,j)\in\mV{s+1}$, by Lemma \ref{lem: forward_argu} it follows that $j=p_m(F_s)$ for some intermediate stage $m$ with $s<m<t$. This proves the set inclusion in \eqref{eq:Ds_bound}.
\end{proof}

For an $F$-pair $(s,N)$, $\mV{N}$ is determined by $\mathbf{K}$ and is independent of the failure sequence. The bound on $\mathcal{D}_s$ for this case is therefore given as follows.

\begin{coro}[Bound on the size of $\mathcal{D}_s$]
      \label{coro: FV} 
    Let $(s,N)$ be an $F$-pair. If $F_s\notin \mathbf{K}$, then 
    \begin{equation} \mathcal{D}_s\subseteq \{v_{s+1}(F_s,j)\mid  j\in\sP{s+1,N-1}(F_s)\}.
    \label{eq:Ds_bound2}
    \end{equation}
    Otherwise, \begin{equation} 
    \mathcal{D}_s=\{v_{s+1}(F_s,j)\mid j\in[\alpha]\}. \label{eq:Ds_bound3}
    \end{equation}
\end{coro}  

\begin{proof}
    Recall that $\mV{N}=\{v_N(i,j)\mid i\in\mathbf{K}, j\in[\alpha]\}$ is dependent on the choice of $\mathbf{K}$.
    \begin{itemize}
        \item  Suppose $F_s\notin \mathbf{K}$.

    By the definition, $v_N(F_s,j)\notin\mV{N}$ for every $j\in[\alpha]$. Since $F_s\notin\sF{s+1,N-1}$ as $(s,N)$ is an $F$-pair, by Lemma \ref{lem: forward_argu}, $v_{s+1}(F_s,j)\in\mV{s+1}$ only if $j=p_m(F_s)$ for some intermediate stage $m$. Therefore, \eqref{eq:Ds_bound2} holds.
    
        \item  Suppose $F_s\in\mathbf{K}$.

        For every $j\in[\alpha]$, $v_N(F_s,j)\in\mV{N}$. Since $(s,N)$ is an $F$-pair, $F_s$ is not the failed node of any intermediate stage $m$. The first part of the backward construction stepping back from stage $N$ to $s+1$ guarantees that $v_m(F_s,j)\in\mV{m}$ for every $j\in[\alpha]$. In particular, we obtain the equality in~\eqref{eq:Ds_bound3}.

\end{itemize}
\end{proof}

We next characterize $\mathbf{N}_s$.

\begin{lemma}[Characterization of $\mathbf{N}_s$]
       \label{lem: AN} 
        Let $(s,t)$ be an $F$-pair with $t\leq N$. A node $i$ belongs to $\mathbf{N}_s$ if either of the following conditions holds:
        \begin{enumerate}
            \item   $i=F_m$ for some intermediate stage $m$ with $F_m\notin \sF{s,m-1}$ and $p_s(F_m)\notin\sP{s+1,m-1}(F_m)$.
            \item  $i\notin \sF{s,t-1}$, $v_t(i,p_s(i))\notin \mV{t}$ and $p_s(i)\notin \sP{s+1,t-1}(i)$.
        \end{enumerate}
\end{lemma}
    
\begin{proof}
    We establish the availability of each type of node separately. 
   
   \begin{itemize}
       \item For the first case, since $F_m$ is the failed node at stage $m$, we have $v_m(F_m,j)\notin\mV{m}$ for every $j\in[\alpha]$. Suppose for contradiction that $v_{s+1}(F_m,p_s(F_m))\in\mV{s+1}$. Since $F_m\notin\sF{s,m-1}$, by Lemma \ref{lem: forward_argu}, $p_s(F_m)=p_c(F_m)$ for some $s+1\leq c<m$, i.e., $p_s(F_m)\in\sP{s+1,m-1}(F_m)$, contradicting the assumption. Therefore, $v_{s+1}(F_m,p_s(F_m))\notin\mV{s+1}$ and $F_m\in \mathbf{N}_s$ by the definition.

       \item For the second case, since $i\notin\sF{s,t-1}$ and $v_t(i,p_s(i))\notin \mV{t}$, Lemma \ref{lem: forward_argu} shows that $v_{s+1}(i,p_s(i))\in\mV{s+1}$ only if $p_s(i)=p_m(i)$ for some $s+1\leq m<t$, i.e., only if $p_s(i)\in \sP{s+1,t-1}(i)$, contradicting the assumption. Therefore, $v_{s+1}(i,p_s(i))\notin\mV{s+1}$, and since $i\neq F_s$, we conclude $i\in\mathbf{N}_s$.
   \end{itemize}
\end{proof}

Under the admissibility condition, case 1 of Lemma \ref{lem: AN} simplifies to the following corollary.

\begin{coro}
    \label{coro: AN_admis}
    Let $(s,t)$ be an $F$-pair. If $m$ is an intermediate stage satisfying $F_m \notin \sF{s,m-1}$ and $|\sF{s,m-1}| \leq \alpha$, then $F_m \in \mathbf{N}_s$.
\end{coro}

\begin{proof}
    Since $|\sF{s,m-1}| \leq \alpha$ and $F_m \notin \sF{s,m-1}$, by admissibility, $p_s(F_m) = p_c(F_m)$ for some $s < c < m$ if and only if $F_s = F_c$. Since $(s,t)$ is an $F$-pair and $m$ is an intermediate stage, $F_s \notin \sF{s+1,m-1}$, which implies no such $c$ exists. Therefore $p_s(F_m) \notin \sP{s+1,m-1}(F_m)$, and Lemma~\ref{lem: AN} gives $F_m \in \mathbf{N}_s$.
\end{proof}

We then prove that the truncated admissible signal flow graph at stage $N$ satisfies the recovery linking condition at stage $N$. By the definition of the active vertices, this is equivalent to the following theorem.

\begin{theorem}
\label{thm: main}
    Suppose the signal flow graph is admissible and truncated at stage $N$. For any $\mathbf{K}=\{i_1,\dots, i_k\}$, with $\mV{N}=\{v_N(i,j)\mid i\in \mathbf{K},j\in[\alpha]\}$, the maximum of $|\mV{0}|$ is at least $B$. Furthermore, a value of at least $B$ is achieved by our recursive construction of $\mV{N-1}$, $\mV{N-2},\ldots, \mV{0}$.
\end{theorem}

To prove Theorem \ref{thm: main}, we partition all stages into two classes. 

\begin{definition} Given a signal flow graph truncated at stage $N$,
a stage $s$ is called {\em extendable} if the $F$-pair $(s,t)$ satisfies $t<N$, and {\em terminal} if $t=N$. 
\end{definition}

We will show that a decrease in $|\mV{s}|$ relative to $|\mV{s+1}|$ occurs only at terminal stages at which $F_s\in\mathbf{K}$. Theorems \ref{thm: invari} and \ref{thm: last} establish these two cases, respectively.
\begin{theorem}
    \label{thm: invari}
  For an extendable stage $s$, given $\mV{s+1}$, the maximum of $|\mV{s}|$ is $|\mV{s+1}|$.
\end{theorem}

\begin{proof}
Let $(s,t)$ be the $F$-pair. Then $F_m\neq F_s$ for any intermediate stage $s<m<t$. By Lemma \ref{lem: Act_bound}, it suffices to prove $|\mathbf{N}_s|\geq |\mathcal{D}_s|$. We handle the two quantities via Lemmas~\ref{lem: FV} and~\ref{lem: AN}, respectively. We consider two cases according to $|\sF{s+1,t-1}|$. 

\begin{itemize}
    \item Assume $|\sF{s+1,t-1}|< \alpha$. Since $(s,t)$ is an $F$-pair, $F_s\notin \sF{s+1,t-1}$. Therefore, by Corollary \ref{coro: distinct_choice}, $|\sP{s+1,t-1}(F_s)|=|\sF{s+1,t-1}|$. 

For any $i\in\sF{s+1,t-1}$, let $m$ be the smallest intermediate stage with $F_m=i$. Then we have $F_m\notin \sF{s,m-1}$ and $|\sF{s+1,m-1}|< |\sF{s+1,t-1}|<\alpha$ and thus $|\sF{s,m-1}|\leq \alpha$. Corollary \ref{coro: AN_admis} then gives $F_m\in\mathbf{N}_s$. It follows that $|\mathbf{N}_s|\geq |\sF{s+1,t-1}|$.

By Lemma \ref{lem: FV}, $\mathcal{D}_s\subseteq \{v_{s+1}(F_s,p_m(F_s))\mid m=s+1,\dots,t-1\}$. The right-hand set has cardinality exactly $|\sP{s+1,t-1}(F_s)|$, i.e., $|\mathcal{D}_s|\leq |\sP{s+1,t-1}(F_s)|$. 

Combining the two bounds yields $$|\mathcal{D}_s|\leq|\sP{s+1,t-1}(F_s)|=|\sF{s+1,t-1}|\leq |\mathbf{N}_s|.$$

\item   If $|\sF{s+1,t-1}|\geq \alpha$, there always exists an intermediate stage $c$ such that $|\sF{s+1,c}|=\alpha$. For any $i\in\sF{s+1,c}$, let $m$ be the smallest stage with $s<m\leq c$ at which $F_m=i$. Then we have $F_m\notin\sF{s,m-1}$ and $|\sF{s,m-1}|\leq \alpha$. By Corollary \ref{coro: AN_admis}, $F_m\in\mathbf{N}_s$ and therefore $|\mathbf{N}_s|\geq |\sF{s+1,c}|=\alpha$.

Since the bound $|\mathcal{D}_s|\leq \alpha$ holds trivially by the definition of failed vertices, we conclude $|\mathbf{N}_s|\geq \alpha\geq |\mathcal{D}_s|$.
\end{itemize}
\end{proof}

\begin{theorem}
   \label{thm: last}
   For a terminal stage $s$, given $\mV{s+1}$, if $F_s\notin\mathbf{K}$, then the maximum of $|\mV{s}|$ is $|\mV{s+1}|$. If $F_s\in\mathbf{K}$, then the maximum of $|\mV{s}|$ is at least
$$
|\mV{s+1}|-\max(0,\alpha-(d-k+|\sF{s,N-1}\cap \mathbf{K}|)).
$$
\end{theorem}

\begin{proof}
    Since stage $s$ is a terminal stage, $(s,N)$ is an $F$-pair, so $F_m\neq F_s$ for any $s<m<N$. By Lemma \ref{lem: Act_bound}, it suffices to bound $|\mathcal{D}_s|$ and $|\mathbf{N}_s|$, and then combine them to determine $|\mV{s}|$. We bound the two quantities separately.

    \textbf{Bounding $|\mathbf{N}_s|$.} We consider the following two cases according to $|\sF{s+1,N-1}|$.
     \begin{itemize}
        \item  Suppose $|\sF{s+1,N-1}|< \alpha$. For any $i\in\sF{s+1,N-1}$, the same argument as in case 1 of the proof of Theorem \ref{thm: invari}, with $t$ replaced by $N$, gives $F_m\in\mathbf{N}_s$. This accounts for $|\sF{s+1,N-1}|$ nodes.

        Moreover, consider a node $i$ with $i\notin\sF{s,N-1}$ and $i\notin\mathbf{K}$. Since $i\notin\mathbf{K}$, by the definition of active vertices we have $v_N(i,p_s(i))\notin\mV{N}$. 
Since $(s,N)$ is an $F$-pair, $F_s\notin \sF{s+1,N-1}$. Since $|\sF{s+1,N-1}|< \alpha$, it follows that $|\sF{s,N-1}|\leq \alpha$. Therefore by applying admissibility, since $i\notin\sF{s,N-1}$, $p_s(i)\in\sP{s+1,N-1}(i)$ if and only if $F_s=F_m$ for some intermediate stage $m$, contradicting $F_s\notin\sF{s+1,N-1}$. Therefore, $p_s(i)\notin \sP{s+1,N-1}(i)$. By Lemma \ref{lem: AN},  such a node $i$ is available at stage $s$. The number of such nodes is $n-|\mathbf{K}\cup\sF{s,N-1}|$.

Since the above two sets of nodes are disjoint and $F_s \notin \sF{s+1,N-1}$,
\begin{align*}
|\mathbf{N}_s|
  &\geq |\sF{s+1,N-1}| + n - |\mathbf{K} \cup \sF{s,N-1}| \\
  &= |\sF{s+1,N-1}| + n - k - |\sF{s,N-1}| + |\mathbf{K} \cap \sF{s,N-1}| \\
  &= |\sF{s+1,N-1}| + n - k - (|\sF{s+1,N-1}|+1) + |\sF{s,N-1} \cap \mathbf{K}| \\
  &= d - k + |\sF{s,N-1} \cap \mathbf{K}|.
\end{align*}
        \item If $|\sF{s+1,N-1}|\geq \alpha$, the same argument as in case 2 of the proof of Theorem \ref{thm: invari}, with $t$ replaced by $N$, gives $|\mathbf{N}_s|\geq \alpha$.
    \end{itemize}

    \textbf{Bounding $|\mathcal{D}_s|$.} By Corollary \ref{coro: FV},
    \begin{itemize}
        \item If $F_s\notin\mathbf{K}$, $\mathcal{D}_s\subseteq \{v_{s+1}(F_s,p_m(F_s))\mid m=s+1,\dots,N-1\}$. Combining the trivial bound that $|\mathcal{D}_s|\leq \alpha$, we have $|\mathcal{D}_s|\leq \min (|\sP{s+1,N-1}(F_s)|,\alpha)$. 
        \item If $F_s\in \mathbf{K}$, $\mathcal{D}_s=\{v_{s+1}(F_s,j)\mid j\in[\alpha]\}$, i.e., $|\mathcal{D}_s|=\alpha$. 
    \end{itemize}

    \textbf{Determining the maximum of $|\mV{s}|$.} By Lemma \ref{lem: Act_bound}, the above discussion implies the following four cases:
    \begin{itemize}
        \item $F_s\notin \mathbf{K}$ and $|\sF{s+1,N-1}|< \alpha$. Since $F_m\neq F_s$ for every intermediate stage $m$, by Corollary \ref{coro: distinct_choice}, $|\sP{s+1,N-1}(F_s)|=|\sF{s+1,N-1}|< \alpha$. It follows that $$|\mathbf{N}_s|\geq  |\sF{s+1,N-1}|=|\sP{s+1,N-1}(F_s)|\geq |\mathcal{D}_s|,$$
and thus the maximum of $|\mV{s}|$ is $|\mV{s+1}|$.
\item     $F_s\notin \mathbf{K}$ and $|\sF{s+1,N-1}|\geq \alpha$. Simply combining the two bounds of failed vertices and available nodes yields$$|\mathbf{N}_s|\geq \alpha\geq |\mathcal{D}_s|,$$ and thus the maximum of $|\mV{s}|$ is $|\mV{s+1}|$.
\item $F_s\in\mathbf{K}$ and $|\sF{s+1,N-1}|< \alpha$. For this case, $|\mathcal{D}_s|=\alpha$ and $|\mathbf{N}_s|\geq d-k+|\sF{s,N-1}\cap \mathbf{K}|$. Applying Lemma \ref{lem: Act_bound}, the maximum of $|\mV{s}|$ is therefore at least $$|\mV{s+1}|-\max(0,\alpha-(d-k+|\sF{s,N-1}\cap \mathbf{K}|)).$$
\item $F_s\in \mathbf{K}$ and $|\sF{s+1,N-1}|\geq\alpha$. Since $(s,N)$ is an $F$-pair, $F_s\notin \sF{s+1,N-1}$ and thus $|\sF{s,N-1}|\geq \alpha+1$. Since $\alpha\geq n-k$, we then have $|\sF{s,N-1}\cap \mathbf{K}|\geq \alpha+1-(n-k)$, which implies that $$\alpha-(d-k+|\sF{s,N-1}\cap \mathbf{K}|)\leq 0.$$ Therefore, the maximum of $|\mV{s}|$ in the statement is exactly $|\mV{s+1}|$. Since $|\mathcal{D}_s|=\alpha$ and $|\mathbf{N}_s|\geq \alpha$ in this case, this maximum is attained.
\end{itemize}
\end{proof}

With Theorems~\ref{thm: invari} and \ref{thm: last} in place, we now prove Theorem \ref{thm: main}.

\begin{proof}[Proof of Theorem \ref{thm: main}]
For any terminal stage $s$, if $F_s\notin\mathbf{K}$, then $|\sF{s,N-1}\cap \mathbf{K}|=|\sF{s+1,N-1}\cap \mathbf{K}|$. If $F_s\in\mathbf{K}$, the quantity increases by $1$, since $F_s\notin \sF{s+1,N-1}$ by the definition of terminal stage. Suppose there are $r$ terminal stages $s$ with $F_s\in\mathbf{K}$. Each such stage contributes the distinct node $F_s$ to $\sF{s,N-1}\cap\mathbf{K}$, and no node leaves this set as $s$ decreases; the values of $|\sF{s,N-1}\cap \mathbf{K}|$ are therefore exactly $1,2,\dots,r$.

By Theorem \ref{thm: invari}, the maximum size of active vertices is preserved at every extendable stage. By Theorem \ref{thm: last}, at each terminal stage $s$ the maximum of $|\mV{s}|$ decreases relative to $|\mV{s+1}|$ by at most $\max(0,\alpha-(d-k+|\sF{s,N-1}\cap \mathbf{K}|))$. Since $\alpha=n-\ell=d+1-\ell$, we have $\alpha-(d-k+|\sF{s,N-1}\cap\mathbf{K}|)=k-\ell+1-|\sF{s,N-1}\cap\mathbf{K}|$.

Summing over all stages, the maximum of $|\mV{0}|$ is at least $$|\mV{N}|- \sum_{x=1}^r \max (0,k-\ell+1-x)\geq |\mV{N}|-\sum_{x=1}^{k-\ell} (k-\ell+1-x)=B,$$
where $|\mV{N}|=k\alpha$ and $B=k\alpha-(1+2+\dots+(k-\ell))$.
\end{proof}

\subsection{Admissibility implies the repair linking condition}\label{subsec:repair}

This subsection establishes that the admissible signal flow graph satisfies the repair linking condition at every stage.
By the definition of the repair linking condition, the main theorem of this subsection is given as follows.

\begin{theorem}
   In the admissible signal flow graph, for any stage $t$, $\{v_t(i,p_t(i))\mid i\neq F_t\}$ can be linked into $\sV_{-1}$, whereas $\{v_t(i,p_t(i))\mid i\neq F_t\}\cup \{v_{t+1}(F_t,j)\}$ cannot be linked into $\sV_{-1}$ for every $j\in[\alpha]$.
    \label{thm: IO_linking}
\end{theorem}

The proof is divided into two parts: Theorems~\ref{thm:positive_linking} and \ref{thm:negative_linking} in this subsection.

We first prove the positive linking statement using a backward-shifting induction. Starting from the repair vertex set at a reference stage $t$, i.e., $\{v_t(i,p_t(i))\mid i\neq F_t\}$, we link it within a finite admissible window onto an earlier target set. The corollary then shows that, under the condition that $p_c(i)=p_t(i)$ for every $i\notin\sF{c,t}$, the repair vertex set at stage $t$ can be linked onto the repair vertex set at stage $c$. The induction hypothesis can therefore be applied. 

The following lemma provides the general backward shift within a finite admissible window. 

\begin{lemma}\label{lem: IO_linking_general}
Fix stages $b<t$ such that $F_t\notin \sF{b,t-1}$ and $|\sF{b,t}|\leq \alpha$, and fix a node $x\notin\sF{b,t-1}$.
Then the set $\{v_t(i,p_t(i))\mid i\neq F_t\}$ can be linked onto $$\{v_b(i,p_t(i))\mid i\notin \sF{b,t}\}\cup\{v_b(x,j)\mid j\in \sP{b,t-1}(x)\}.$$
\end{lemma}

\begin{proof}
For each $i\neq F_t$, we construct a path $P_i$ from $v_t(i,p_t(i))$ to the target set in the statement. Since $F_t\notin\sF{b,t-1}$, the two cases $i\notin\sF{b,t}$ and $i\in\sF{b,t-1}$ partition all nodes $i\neq F_t$.
\begin{itemize}
\item Suppose $i\notin\sF{b,t}$. Since node $i$ does not fail from stage $b$ to stage $t$, define
$$
P_i=\bigl(v_t(i,p_t(i)),\ldots,v_b(i,p_t(i))\bigr),
$$
where every consecutive pair is connected by a carry-forward edge. This path terminates at $v_b(i,p_t(i))$, which belongs to the first part of the target set.
\item Suppose $i\in\sF{b,t-1}$. Let $b\leq m\leq t-1$ be the largest stage such that $F_m=i$. Then $F_m\notin\sF{m+1,t}$, and since $x\notin\sF{b,t-1}$, the repair edge from $v_m(x,p_m(x))$ to $v_{m+1}(i,p_t(i))$ exists. Define
$$
P_i=\bigl(v_t(i,p_t(i)),\ldots,v_{m+1}(i,p_t(i)),v_m(x,p_m(x)),\ldots,v_b(x,p_m(x))\bigr),
$$
where the first and last segments follow carry-forward edges and the transition uses the above repair edge. This path terminates at $v_b(x,p_m(x))$, which belongs to the second part of the target set.

\end{itemize}

We next verify that the target vertices cover the whole target set. The first case clearly covers $\{v_b(i,p_t(i))\mid i\notin\sF{b,t}\}$. For the second case, since $x\notin\sF{b,t-1}$ and $|\sF{b,t-1}|\leq|\sF{b,t}|\leq\alpha$, by Corollary \ref{coro: distinct_choice} $|\sP{b,t-1}(x)|=|\sF{b,t-1}|$, and the values $p_m(x)$ corresponding to the distinct nodes in $\sF{b,t-1}$ are pairwise distinct. Therefore, as $i$ ranges over $\sF{b,t-1}$, the endpoints in the second case cover exactly $\{v_b(x,j)\mid j\in\sP{b,t-1}(x)\}$.

    It remains to check that the paths are mutually disjoint. Paths in the first case use only nodes outside $\sF{b,t}$, while paths in the second case first use a node in $\sF{b,t-1}$ and then the fixed node $x$. Since $x\notin\sF{b,t-1}$, the corresponding path in the first case is $P_x$ when $x\neq F_t$. We then have $p_t(x)\neq p_m(x)$ for any $b\leq m\leq t-1$ by admissibility. Paths from the first case are therefore disjoint from paths from the second case. 

    The paths in the first case are mutually disjoint. For paths in the second case corresponding to distinct nodes, their first segments lie on different nodes, and their last segments lie on node $x$ with distinct packet indices. Therefore these paths are also mutually disjoint. Thus the constructed paths form the desired linking.
\end{proof}

The following corollary applies Lemma~\ref{lem: IO_linking_general} with $b=c+1$ and $x=F_c$. Under the assumption that $p_c(i)=p_t(i)$ for all nodes $i\notin\sF{c,t}$, the resulting target is exactly the repair vertex set at stage $c$.

\begin{coro}\label{coro: IO_linking_original}
Fix stages $c<t$ such that $F_t\notin\sF{c+1,t-1}$, $F_c\notin\sF{c+1,t-1}$, and $|\sF{c+1,t}|\leq \alpha$. If, in addition, 
$p_c(i)=p_t(i)$ for every $i\notin\sF{c,t}$, then the set $\{v_t(i,p_t(i))\mid i\neq F_t\}$ can be linked onto $\{v_c(i,p_c(i))\mid i\neq F_c\}$.
\end{coro}

\begin{proof}
Apply Lemma~\ref{lem: IO_linking_general} at stage $c+1$ with $x=F_c$. This is valid because $F_t\notin\sF{c+1,t-1}$, $|\sF{c+1,t}|\leq\alpha$, and $F_c\notin\sF{c+1,t-1}$. Therefore, $\{v_t(i,p_t(i))\mid i\neq F_t\}$ can be linked onto the lemma target set
$$
\{v_{c+1}(i,p_t(i))\mid i\notin\sF{c+1,t}\}\cup\{v_{c+1}(F_c,j)\mid j\in\sP{c+1,t-1}(F_c)\}.
$$
It remains to extend this linking from stage $c+1$ to the desired target set at stage $c$ in the statement. 

First consider the vertices in the lemma target set with node index outside $\sF{c,t}$. These vertices are exactly $\{v_{c+1}(i,p_t(i))\mid i\notin\sF{c,t}\}$, and each of them is linked by a carry-forward edge to the corresponding vertex in $\{v_c(i,p_t(i))\mid i\notin\sF{c,t}\}$.

To handle the remaining vertices in the lemma target set, since all these vertices are on node $F_c$, we consider the following two cases.

\begin{itemize}
    \item If $F_c\neq F_t$, then $F_c\notin\sF{c+1,t}$, so the vertex $v_{c+1}(F_c,p_t(F_c))$ comes from the first part of the lemma target set. Together with the second part, the remaining vertices on node $F_c$ are precisely
$$
\{v_{c+1}(F_c,j)\mid j\in\sP{c+1,t}(F_c)\}.
$$
    Since $F_c\notin\sF{c+1,t}$ and $|\sF{c+1,t}|\leq\alpha$, Corollary \ref{coro: distinct_choice} gives $|\sP{c+1,t}(F_c)|=|\sF{c+1,t}|$. Moreover, the repair edges at stage $c$ connect the vertices $v_{c+1}(F_c,j)$ to $v_c(i,p_c(i))$ with $i\in\sF{c+1,t}$, which gives a linking from this set onto $\{v_c(i,p_c(i))\mid i\in\sF{c+1,t}\}$.
\item If $F_c=F_t$, then $p_t(F_c)$ is not defined. In this case, the remaining vertices of the lemma target set on node $F_c$ are precisely
$$
\{v_{c+1}(F_c,j)\mid j\in\sP{c+1,t-1}(F_c)\}.
$$
    Since $F_c\notin\sF{c+1,t-1}$ and $|\sF{c+1,t-1}|\leq\alpha$, Corollary \ref{coro: distinct_choice} gives $|\sP{c+1,t-1}(F_c)|=|\sF{c+1,t-1}|$. The repair edges at stage $c$ therefore give a linking of the set onto $\{v_c(i,p_c(i))\mid i\in\sF{c+1,t-1}\}$.
\end{itemize}

Since the two parts of the linking are clearly disjoint, combining the carry-forward part and the repair part gives a linking of the repair vertex set at stage $t$ onto $\{v_c(i,p_t(i))\mid i\notin\sF{c,t}\}\cup\{v_c(i,p_c(i))\mid i\in\sF{c,t}\setminus\{F_c\}\}$. By the assumption that $p_t(i)=p_c(i)$ for every $i\notin \sF{c,t}$, the final target set is therefore $\{v_c(i,p_c(i))\mid i\neq F_c\}$.
\end{proof}

    We now prove the positive linking statement by induction on $t$, using the lemma and the corollary to shift the repair vertex set to an earlier stage where the induction hypothesis or the initialization applies.

\begin{theorem}
    For any stage $0\leq t\leq N-1$, $\{v_t(i,p_t(i))\mid i\neq F_t\}$ can be linked into $\sV_{-1}$. \label{thm:positive_linking}
\end{theorem}

\begin{proof}
    We use induction to prove the theorem.

    For $t=0$, note that $|\{v_0(i,p_0(i))\mid i \neq F_0\}|=n-1$; recall that $n-1\leq B$. According to the initialization of stage $0$, $\{v_0(i,p_0(i))\mid i\neq F_0\}$ can therefore be linked into $\sV_{-1}$. 

 For $t>0$, suppose the statement holds for stages $0,1,\dots,t-1$. Consider the following cases.

 \begin{itemize}
     \item Suppose $(s,t)$ is an $F$-pair, with $|\sF{s+1,t}|\leq \alpha$. Therefore, $F_s=F_t$ and $F_s\notin\sF{s+1,t-1}$. By admissibility, it follows that $p_s(i)=p_t(i)$ for any $i\notin \sF{s,t}$. 

     By applying Corollary \ref{coro: IO_linking_original} at stage $s$, $\{v_t(i,p_t(i))\mid i\neq F_t\}$ can be linked onto $$\{v_s(i,p_s(i))\mid i\neq F_s\}.$$ The induction assumption implies that the set can be further linked into $\sV_{-1}$.

\item Suppose there does not exist a stage $s$ such that $(s,t)$ is an $F$-pair, and $|\sF{0,t}|\leq \alpha$. Then $F_t\notin\sF{0,t-1}$, and there always exists a node $x\notin\sF{0,t}$. Applying Lemma \ref{lem: IO_linking_general} at stage $0$ then yields that $\{v_t(i,p_t(i))\mid i\neq F_t\}$ can be linked onto $$\{v_0(i,p_t(i))\mid i\notin\sF{0,t}\}\cup \{v_0(x,j)\mid j\in\sP{0,t-1}(x)\}.$$  Since this set is a vertex set at stage $0$ of size $n-1\leq B$, by the initialization of stage $0$, it can be linked into $\sV_{-1}$.

\item It remains to consider the cases in which the relevant window ending at $t$ contains more than $\alpha$ distinct failed nodes, namely either the $F$-pair $(s,t)$ exists with $|\sF{s+1,t}|>\alpha$, or no such $F$-pair exists and $|\sF{0,t}|>\alpha$. Since $|\sF{m,t}|$ increases by at most one as $m$ decreases and eventually exceeds $\alpha$, there exists a stage $c<t$ with $|\sF{c+1,t}|=\alpha$ and $F_c\notin\sF{c+1,t}$. It follows that $c>s$ if an $F$-pair $(s,t)$ exists, and $c\geq 0$ otherwise. Therefore, in both cases $F_t\notin\sF{c+1,t-1}$ and $|\sF{c+1,t-1}|=\alpha-1$.

    For every $i\notin\sF{c,t}$, admissibility gives $|\sP{c+1,t-1}(i)|=\alpha-1$ and $p_t(i)\notin\sP{c+1,t-1}(i)$. Moreover, since $F_c\notin\sF{c+1,t-1}$, a similar argument yields $p_c(i)\notin\sP{c+1,t-1}(i)$. Both $p_t(i)$ and $p_c(i)$ are therefore the unique element of $[\alpha]\setminus\sP{c+1,t-1}(i)$, i.e., $p_c(i)=p_t(i)$.

    We can then apply Corollary \ref{coro: IO_linking_original} at stage $c$, and therefore $\{v_t(i,p_t(i))\mid i\neq F_t\}$ can be linked onto $$\{v_c(i,p_c(i))\mid i\neq F_c\}.$$
Since $c<t$, by the induction assumption, it can be further linked into $\sV_{-1}$.
\end{itemize}     
\end{proof}

It remains to verify the negative part of the repair linking condition.

\begin{theorem}
        For any $t$ and $j$, no linking of $\{v_t(i,p_t(i))\mid i\neq F_t\}\cup \{v_{t+1}(F_t,j)\}$ into $\sV_{-1}$ exists. \label{thm:negative_linking}
\end{theorem}

\begin{proof}
Suppose for contradiction that such a linking exists. Consider the path starting from the newcomer vertex $v_{t+1}(F_t,j)$. By the construction of the signal flow graph, only repair edges are incident to this vertex. Therefore, to reach $\sV_{-1}$, the path must contain a vertex $v_t(i,p_t(i))$ with $i\neq F_t$. 

However, every $v_t(i,p_t(i))$ with $i\neq F_t$ is a source vertex of the required linking. The newcomer path therefore shares a vertex with the path starting from $v_t(i,p_t(i))$, contradicting the mutual disjointness of the linking. It follows that such a linking does not exist.
\end{proof}

By combining Theorem~\ref{thm:positive_linking} and Theorem~\ref{thm:negative_linking}, we obtain Theorem~\ref{thm: IO_linking}. Thus the admissible signal flow graph satisfies the repair linking condition at every stage.

\subsection{Algorithmic Construction} \label{subsec: alg}

The preceding two subsections establish that an admissible signal flow graph satisfies both linking conditions. This subsection presents an explicit algorithm for constructing such a graph. For each node and stage, the algorithm determines the choice function by examining a window containing at most $\alpha$ distinct failed nodes. We then prove that the choice functions so produced yield an admissible signal flow graph.

\begin{algorithm}[H]
	\renewcommand{\algorithmicrequire}{\textbf{Input:}}
	\renewcommand{\algorithmicensure}{\textbf{Output:}}
	\caption{Finite-window computation of the choice function}
	\label{alg1}
	\begin{algorithmic}
		\REQUIRE Node $i$, stage $t$.
		\ENSURE The choice function $p_t(i)$, if $i\neq F_t$.
		\IF{$i=F_t$}
			\STATE $p_t(i)$ is undefined.
		\ELSE
			\STATE $c\leftarrow t-1$.
			\WHILE{$c\geq 0$, $F_c\notin\{F_t,i\}$, and $|\sF{c,t-1}|<\alpha$}
				\STATE $c\leftarrow c-1$.
			\ENDWHILE
			\IF{$c=-1$}
				\STATE Choose $p_t(i)\in[\alpha]\setminus\sP{0,t-1}(i)$.
			\ELSIF{$F_c=F_t$}
				\STATE $p_t(i)\leftarrow p_c(i)$.
			\ELSIF{$F_c=i$}
				\STATE Choose $p_t(i)\in[\alpha]\setminus\sP{c+1,t-1}(i)$.
			\ELSE
				\STATE $p_t(i)\leftarrow p_c(i)$.
			\ENDIF
		\ENDIF
	\end{algorithmic}
\end{algorithm}

\begin{remark}[Finite-memory implementation]
Although Algorithm~\ref{alg1} is presented as a backward search, it suffices to maintain an ordered sequence of length $\alpha$ consisting of the most recently distinct failed nodes and the values of the associated choice functions. When a failure occurs, the corresponding record is updated and the oldest record is discarded if necessary. 

The stopping stage and the choice function required by Algorithm~\ref{alg1} can then be recovered
from this maintained sequence. Thus, repeated failures may cause the corresponding interval in the original failure sequence to span many stages, but the length of the sequence is finite and independent of the stage index $t$.
\end{remark}

The choice functions of each node are determined independently at each stage. We now prove that the signal flow graph generated by the algorithm satisfies the admissibility condition introduced at the beginning of this section.

\begin{theorem}
    The signal flow graph generated by Algorithm \ref{alg1} is admissible.
    \label{thm: alg}
\end{theorem}
\begin{proof}
    See Appendix \ref{apd: 1}.
\end{proof}

By this theorem, Algorithm~\ref{alg1} generates an admissible signal flow graph for any failure sequence. Since an admissible signal flow graph satisfies the two required linking conditions at every stage, this completes the required combinatorial construction.

\begin{example}
Consider parameters $n=5$, $k=3$, $\alpha=3$, $d=4$, $\beta=1$, $N=9$, with node set $[5]$ and $[\alpha]=\{1,2,3\}$. We apply Algorithm~\ref{alg1} to the failure sequence
\[
(F_0,F_1,\dots,F_8)=(1,2,3,4,4,1,3,5,2),
\]
in which node $1$ fails at stages $0$ and $5$, and node $3$ fails at stages $2$ and $6$. The resulting choice functions are recorded in Table~\ref{tab:example}.

\begin{table}[h]
\centering
\caption{Choice functions $p_t(i)$ produced by Algorithm~\ref{alg1} for the failure sequence $(1,2,3,4,4,1,3,5,2)$.}
\label{tab:example}
\begin{tabular}{ccccccc}
\hline
$t$ & $F_t$ & $p_t(1)$ & $p_t(2)$ & $p_t(3)$ & $p_t(4)$ & $p_t(5)$ \\
\hline
0 & 1 & $-$ & 1 & 1 & 1 & 1 \\
1 & 2 & 1 & $-$ & 2 & 2 & 2 \\
2 & 3 & 2 & 1 & $-$ & 3 & 3 \\
3 & 4 & 3 & 2 & 1 & $-$ & 1 \\
4 & 4 & 3 & 2 & 1 & $-$ & 1 \\
5 & 1 & $-$ & 3 & 2 & 1 & 2 \\
6 & 3 & 1 & 1 & $-$ & 2 & 3 \\
7 & 5 & 2 & 2 & 1 & 3 & $-$ \\
8 & 2 & 3 & $-$ & 2 & 1 & 1 \\
\hline
\end{tabular}
\end{table}

We verify admissibility on several representative windows.

\textbf{Node 2, window $[2,6]$.} Since $2\notin\sF{2,6}=\{3,4,1\}$ and $|\sF{2,6}|=3=\alpha$, both directions of the admissibility condition must hold for every stage pair in this window. The repeated failure $F_2=F_6=3$ requires $p_2(2)=p_6(2)$, which the table confirms with $p_2(2)=p_6(2)=1$. The remaining distinct failures in the window yield distinct choice functions, and accordingly $p_2(2)=1$, $p_3(2)=p_4(2)=2$, $p_5(2)=3$. 

\textbf{Node 4, window $[5,7]$.} Since $4\notin\sF{5,7}=\{1,3,5\}$ and the three failures are distinct, admissibility requires the three choice functions to be pairwise distinct, and indeed $p_5(4)=1$, $p_6(4)=2$, $p_7(4)=3$ exhaust $[\alpha]$.

The two repeated failures behave differently with respect to admissibility. The repetition of node $3$ at stages $2$ and $6$ satisfies $|\sF{2,6}|\leq\alpha$ and the equality $p_2(i)=p_6(i)$ is enforced for every node $i\notin\sF{2,6}$. In contrast, the repetition of node $1$ at stages $0$ and $5$ satisfies $|\sF{0,5}|>\alpha$, so admissibility imposes no constraint between them; for instance, $p_0(2)=1\neq 3=p_5(2)$ causes no violation. The remaining windows are verified similarly, and the signal flow graph generated by Algorithm~\ref{alg1} is therefore admissible, consistent with Theorem~\ref{thm: alg}.
\end{example}

\begin{theorem}
    Suppose $d=n-1$. Fix integers $n,k,\ell$ with $1\leq k\leq n-1,\ell\in\{1,\dots,k\}$, and let $\alpha,\beta=1$ and $B$ be the corresponding parameters. Let $\FF{q}$ be a fixed finite field satisfying $q> \binom{n\alpha}{B}-\binom{(n-1)\alpha}{B}$. Then for any infinite failure sequence $(F_0,F_1,\dots)$, Algorithm~\ref{alg1} together with Theorem~\ref{thm: linear_repre} yields a functional-repair regenerating code with parameters $(n,k,d=n-1,\alpha,\beta=1,B)$ over $\FF{q}$ that is both access-optimal and help-by-transfer, and that maintains the $(n,k)$ recovery property after every repair. 

    Moreover, the field $\FF{q}$ is fixed and each repair is determined by a finite history. The code therefore supports an unbounded number of repairs with per-repair complexity independent of the repair history. Since $\ell$ is arbitrary, every vertex point of the whole tradeoff curve is attainable.
\end{theorem}

\begin{proof}
    By Theorem~\ref{thm: alg}, the constructed signal flow graph is admissible. Theorem~\ref{thm: main} and~\ref{thm: IO_linking} together ensure that the two linking conditions are satisfied at each stage of the constructed graph. Applying Theorem \ref{thm:linking_conditions_to_code} then yields the code satisfying the recovery and repair properties at every stage. 

    Theorem~\ref{thm: linear_repre} and Corollary~\ref{coro: linear_repre} further ensure that the finite field is fixed and that the local encoding coefficients can be selected using finite memory. Algorithm~\ref{alg1} then shows that the choice functions can also be selected using finite memory. The code therefore supports an unbounded number of repairs with per-repair complexity independent of the number of repairs.
\end{proof}

\section{Proof of Theorem \ref{thm: linear_repre} and Corollary \ref{coro: linear_repre}}

\label{sec: linear}

It is important to note that the construction and arguments in this section do not directly rely on the conclusions of Section~\ref{sec:packet_selection}. The gammoids induced by a signal flow graph are linearly representable regardless of whether the signal flow graph is admissible. However, the linear representation only translates the combinatorial independence into linear independence, and does not guarantee that the resulting linear system satisfies the requirements of a functional-repair regenerating code.

For example, the signal flow graph in Fig.~\ref{fig:graph} is admissible, whereas the graph in Fig.~\ref{fig:graph_wrong} is not. Nevertheless, the gammoids associated with both graphs are linearly representable. The difference is that only the code corresponding to Fig.~\ref{fig:graph} satisfies the requirements of a regenerating code. The code associated with the graph in Fig.~\ref{fig:graph_wrong} fails the $(n,k)$ recovery property at stage~$3$.

The goal in the proof of Theorem~\ref{thm: linear_repre} is to show that the vector matroid, and hence the corresponding access-optimal regenerating code, can be constructed incrementally and causally. This is done by establishing an isomorphism between the graphical gammoid and the vector matroid, thereby faithfully representing the abstract gammoid by global encoding vectors. Once this isomorphism is established, the resulting vector matroid inherits the relevant combinatorial properties of the gammoid, such as the $(n,k)$ recovery property.

Recall from~\eqref{eq:transfer1} and~\eqref{eq:transfer2} that the global encoding vectors at stage $t+1$ are given by
\begin{equation}
    \be_{t+1}(i,j)=
    \begin{cases}
    \be_t(i,j), & \text{if } i\neq F_t,\\[1mm]
    \displaystyle\sum_{i'\in[n]\setminus\{F_t\}} b_t(i',j)\be_t(i',p_t(i')), & \text{if } i=F_t,
    \end{cases}
\label{eq:transfer}
\end{equation}
where $j=1,2,\ldots,\alpha$. Thus, the global encoding vectors $\be_{t+1}(i,j)$ are determined by the global encoding vectors at stage $t$ and by the local encoding coefficients $b_t(i,j)$.

We collect the local encoding coefficients into the matrix
\[
\bL_t :=
\begin{pmatrix}
    b_t(1,1)& \cdots & b_t(F_t-1,1)&b_t(F_t+1,1)&\cdots&b_t(n,1)\\
    b_t(1,2)& \cdots & b_t(F_t-1,2)&b_t(F_t+1,2)&\cdots&b_t(n,2)\\
    \vdots&&\vdots&\vdots&&\vdots\\
    b_t(1,\alpha) & \cdots & b_t(F_t-1,\alpha)&b_t(F_t+1,\alpha)&\cdots&b_t(n,\alpha)
\end{pmatrix}.
\]
This matrix has size $\alpha\times(n-1)$. The $j$-th row of $\bL_t$ consists of the coefficients appearing in the summation in~\eqref{eq:transfer}. We note that there is no term of the form $b_t(F_t,j)$.

If the $n-1$ packets transmitted to the newcomer are arranged as the column vector
\[
\mathbf{v}_t :=
\begin{bmatrix}
S_t(1,p_t(1)) \\
\vdots \\
S_t(F_t-1,p_t(F_t-1)) \\
S_t(F_t+1,p_t(F_t+1))\\
\vdots \\
S_t(n,p_t(n))
\end{bmatrix},
\]
then the product $\bL_t\mathbf{v}_t$ is an $\alpha$-dimensional vector whose components are stored in the newcomer.

The entries of $\bL_t$ are the local encoding coefficients that we may choose in order to maintain the recovery linking condition and the repair linking condition defined in Section~\ref{sec: reduction}. The global encoding vectors for the vertices in $\mathcal{V}_t$ have already been determined in the previous stage. We want to choose the local encoding coefficients in $\bL_t$ so as to assign vectors to the vertices in $\mathcal{V}_{t+1}$, and hence to form $\mathcal{G}_{t+1}$, such that the vector matroid formed by $\mathcal{G}_t\cup\mathcal{G}_{t+1}$ is isomorphic to $\mathscr{M}_t$. Equivalently, for any index set $\mathcal{I}$ with
\[
    \mathcal{V}(\mathcal{I})\subseteq \mathcal{V}_t\cup\mathcal{V}_{t+1},
\]
the vectors in $\mathcal{G}(\mathcal{I})$ are linearly independent if and only if the vertices in $\mathcal{V}(\mathcal{I})$ are independent in the matroid $\mathscr{M}_t$. We say such a set $\mathcal{I}$ is of the form $(u,o)$ if $|\mathcal{I}_t|=u $ and $|\mathcal{I}_{t+1}|=o$.

The main goal of this section is to prove the forward direction of Theorem~\ref{thm: linear_repre}. We show that, if the finite field is sufficiently large, then an appropriate matrix $\bL_t$ can be chosen.

Since $\mathscr{M}_{t-1}$ and $\mathscr{M}_t$ are restrictions of the same strict gammoid, their restrictions to $\mathcal{V}_t$ coincide. Hence, the hypothesis of Theorem~\ref{thm: linear_repre} implies that the indexed family $\mathcal{G}_t$ represents the restriction of $\mathscr{M}_t$ to $\mathcal{V}_t$.

\begin{theorem}
    Suppose the size of the finite field  $\mathbb{F}_q$  satisfies
$$
	q> \binom{n\alpha}{B}-\binom{(n-1)\alpha}{B}.$$
    Suppose that the vector matroid on $\mathcal{V}_t$ represented by the indexed family $\mathcal{G}_t$ is isomorphic to the restriction of $\mathscr{M}_t$ to $\mathcal{V}_t$. Then we can find local encoding coefficients $b_t(i,j)$ in $\FF{q}$ such that whenever $\mathcal{V}(\mathcal{I})\in\mathscr{C}_t$, the corresponding vectors in $\mathcal{G}(\mathcal{I})$ are linearly independent over~$\FF{q}$.
    \label{thm: linear_indep}
\end{theorem}

The reverse direction of Theorem~\ref{thm: linear_indep} will be proved in Theorem~\ref{thm: dependent}. The proof of Corollary~\ref{coro: linear_repre} then follows by applying Theorem~\ref{thm: linear_repre} inductively from the initialization at stage $0$.

We introduce some additional notation.

\begin{definition}
Given a matrix $\bL_t$ of local encoding coefficients, we say that $\bL_t$

\begin{enumerate}
\item {\em preserves the independence of $\mathcal{I}$}, for a set $\mathcal{I}$ such that the corresponding $\sV(\mathcal{I})$ is independent in the gammoid $\mathscr{M}_t$, if the corresponding set $\mathcal{G}(\mathcal{I})$ is linearly independent;

\item is {\em $(u,o)$-preserving} if it preserves the independence of every set $\mathcal{I}$ of the form $(u,o)$ such that $\sV(\mathcal{I})$ is independent in $\mathscr{M}_t$;

\item is {\em non-$(u,o)$-preserving} if there exists a set $\mathcal{I}$ of the form $(u,o)$ such that $\sV(\mathcal{I})$ is independent in $\mathscr{M}_t$ while $\bL_t$ does not preserve the independence of $\mathcal{I}$.
\end{enumerate}
\end{definition}

Theorem~\ref{thm: linear_indep} holds if there exists a matrix $\bL_t$ that preserves the independence of every $\mathcal{I}$ corresponding to an independent set in $\mathscr{M}_t$. We first record two reduction lemmas.

\begin{lemma}
To establish that $\bL_t$ preserves the independence of every index set $\mathcal{I}$ such that $\mathcal{V}(\mathcal{I})$ is independent in $\mathscr{M}_t$, it suffices to verify this for those sets $\mathcal{I}$ with size $|\mathcal{I}|\leq B$, such that $\mathcal{I}_{t+1}$ is nonempty and
\[
    \mathcal{I}_{t+1}
    \subseteq
    \{(F_t,j)_{t+1}: j\in[\alpha]\}.
\]
\end{lemma}

\begin{proof}
We consider an index set $\mathcal{I}$ that does not satisfy the stated condition.

First, if $\mathcal{I}_{t+1}$ is empty, then $\mathcal{I}$ contains only indices at stage $t$, i.e.,
\[
    \mathcal{V}(\mathcal{I})\subseteq \mathcal{V}_t.
\]
The corresponding vectors in $\mathcal{G}_t$ are linearly independent regardless of the choice of $\bL_t$, by the assumption in Theorem~\ref{thm: linear_indep} that the representation has already been constructed at stage $t$.

Next, suppose that $(i,j)_{t+1}\in\mathcal{I}$ for some $i\neq F_t$. By~\eqref{eq:transfer}, we have
\[
    \be_{t+1}(i,j)=\be_t(i,j).
\]
In the signal flow graph, the unique edge terminating at $v_{t+1}(i,j)$ is the carry-forward edge from $v_t(i,j)$.

If $(i,j)_t\in\mathcal{I}$, Theorem \ref{thm: graph_property} implies that $\sV(\mathcal{I})$ is dependent in $\mathscr{M}_t$, and $\mathcal{G}(\mathcal{I})$ is linearly dependent since it contains two identical vectors. 

If $(i,j)_t\notin\mathcal{I}$, define
\[
    \mathcal{I}'
    =
    \bigl(\mathcal{I}\setminus\{(i,j)_{t+1}\}\bigr)
    \cup
    \{(i,j)_t\}.
\]
Since $i\neq F_t$, Theorem~\ref{thm: graph_property} implies that $\mathcal{V}(\mathcal{I})$ can be linked into $\mathcal{V}_{-1}$ if and only if $\mathcal{V}(\mathcal{I}')$ can be linked into $\mathcal{V}_{-1}$. Moreover, $\mathcal{G}(\mathcal{I})$ and $\mathcal{G}(\mathcal{I}')$ contain the same vectors. Hence the two sets have the same linear independence property. Therefore, if $\bL_t$ preserves the independence of $\mathcal{I}'$, then it also preserves the independence of $\mathcal{I}$.
\end{proof}

In view of this lemma, throughout the remainder of this section we only need to consider sets $\mathcal{I}$ of the form $(u,o)$ with $1\leq o\leq \alpha$ and
\[
    \mathcal{I}_{t+1}
    \subseteq
    \{(F_t,j)_{t+1}: j\in[\alpha]\}.
\]

The following lemma shows that, for each fixed value of $o$, it suffices to verify the $(u,o)$-preserving property at the largest relevant value of~$u$.

\begin{lemma}
    If $\bL_t$ is $(u,o)$-preserving with $u>1$, then $\mathbf{L}_t$ is also $(u',o)$-preserving for all $u'<u$.
\label{lem: (u,o)-preserving}
\end{lemma}

\begin{proof}
Let $\mathcal{I}'$ be an independent set of the form $(u',o)$, where $u'<u$. Since $u+o\leq B$, we may extend $\mathcal{I}'$ by adding $u-u'$ indices from stage $t$ to obtain an index set $\mathcal{I}$ of the form $(u,o)$ with
\[
    \mathcal{I}'\subseteq \mathcal{I}
    \quad\text{and}\quad
    \mathcal{I}'_{t+1}=\mathcal{I}_{t+1}.
\]
By the augmentation axiom $I$3 of matroids, the additional indices can be chosen so that $\mathcal{V}(\mathcal{I})$ remains independent in $\mathscr{M}_t$.

Since $\bL_t$ is $(u,o)$-preserving, the set $\mathcal{G}(\mathcal{I})$ is linearly independent. Every subset of a linearly independent set is linearly independent; in particular, $\mathcal{G}(\mathcal{I}')$ is linearly independent. Since the argument applies to every independent set $\mathcal{I}'$ of the form $(u',o)$, the matrix $\bL_t$ is $(u',o)$-preserving for all $u'<u$.
\end{proof}

We are now ready to derive the field-size bound in Theorem~\ref{thm: linear_indep}. The proof is a counting argument in the spirit of the Schwartz--Zippel lemma.

\begin{proof}[Proof of Theorem~\ref{thm: linear_indep}]
Since the ambient vector space has dimension $B$, the set $\mathcal{G}(\mathcal{I})$ is necessarily linearly dependent if $|\mathcal{I}|>B$. Likewise, since $\mathcal{V}_{-1}$ has size $B$, the set $\mathcal{V}(\mathcal{I})$ is dependent in $\mathscr{M}_t$ if $|\mathcal{I}|>B$. We therefore assume $u+o\leq B$. By Lemma~\ref{lem: (u,o)-preserving}, for each fixed $o=1,2,\ldots,\alpha$, it suffices to guarantee preservation of independence for $u=B-o$. Hence, it is enough to consider independent sets $\mathcal{V}(\mathcal{I})\in\mathscr{C}_t$ where $\mathcal{I}$ is of the form $(B-o,o)$.

We proceed by considering $o=1,2,\ldots,\alpha$. For each $o$, we bound the number of $\alpha\times(n-1)$ matrices that are non-$(B-o,o)$-preserving but are $(B-o',o')$-preserving for every $o'<o$. Summing these bounds gives an upper bound on the number of matrices that fail to preserve the independence of at least one independent set. Therefore, if the total number of possible matrices $\bL_t$ exceeds this upper bound, then there exists a matrix $\bL_t$ that preserves the independence of every independent set in $\mathscr{M}_t$.

The argument is a generic counting argument and does not depend on the particular indices in $\mathcal{I}$. Without loss of generality, let
\begin{align*}
    \mathcal{I}_t =\{I_1,\ldots,I_{B-o}\},\\
    \mathcal{I}_{t+1}
    &= \{(F_t,1)_{t+1},\ldots,(F_t,o)_{t+1}\}.
\end{align*}
By Theorem~\ref{thm: graph_property}, $\mathcal{V}(\mathcal{I})$ is independent in $\mathscr{M}_t$ if and only if there exists a subset
\[
    \mathcal{Y}
    \subseteq
    \{(i,p_t(i))_t: i\neq F_t\}\setminus \mathcal{I}_t
\]
with $|\mathcal{Y}|=o$ such that
\[
    \mathcal{V}(\mathcal{Y}\cup\mathcal{I}_t)
\]
is independent in $\mathscr{M}_t$.

For notational convenience, we relabel the helper nodes and write
\[
    \mathcal{Y}
    :=
    \{(1,p_t(1))_t,\ldots,(o,p_t(o))_t\}.
\]
Since $\mathcal{V}(\mathcal{Y}\cup\mathcal{I}_t)\subseteq\mathcal{V}_t$, the assumption that $\mathcal{G}_t$ represents the independence relations at stage $t$ implies that the corresponding global encoding vectors are linearly independent.

Fix $o\in\{1,2,\ldots,\alpha\}$ and suppose that $\bL_t$ does not preserve the independence of $\mathcal{I}$, while being $(B-o',o')$-preserving for every $o'<o$.

The $B$ vectors in $\mathcal{G}(\mathcal{I})$, computed using the matrix $\bL_t$, can be arranged as the columns of a $B\times B$ matrix over $\mathbb{F}_q$. By a slight abuse of notation, we write $\det \mathcal{G}(\mathcal{I})$ for the determinant of this matrix. Since $\mathcal{V}(\mathcal{I})$ is independent in $\mathscr{M}_t$, we have
\[
    \bL_t \text{ preserves the independence of }\mathcal{I}
    \quad\Longleftrightarrow\quad
    \det\mathcal{G}(\mathcal{I})\neq 0.
\]

Recall from~\eqref{eq:transfer} that
\[
    \be_{t+1}(F_t,o)
    =
    \sum_{m\in[n]\setminus\{F_t\}}
    b_t(m,o)\be_t(m,p_t(m)).
\]
By the multilinearity of the determinant, expanding $\det\mathcal{G}(\mathcal{I})$ along the column corresponding to $\be_{t+1}(F_t,o)$ gives
\[
    \det \mathcal{G}(\mathcal{I})
    =
    \sum_{m\in[n]\setminus\{F_t\}}
    (\pm)\det \mathcal{G}(\mathcal{I}^{(m)})\, b_t(m,o),
\]
where $\mathcal{I}^{(m)}$ is obtained from $\mathcal{I}$ by replacing $(F_t,o)_{t+1}$ with $(m,p_t(m))_t$. Each $\mathcal{I}^{(m)}$ is of the form $(B-o+1,o-1)$.

For $m=1,\ldots,o$, we have $(m,p_t(m))_t\in\mathcal{Y}$. By Theorem~\ref{thm: graph_property}, it follows that $\mathcal{V}(\mathcal{I}^{(m)})$ is independent in $\mathscr{M}_t$. Since $\bL_t$ is $(B-o+1,o-1)$-preserving by the induction hypothesis, the set $\mathcal{G}(\mathcal{I}^{(m)})$ is linearly independent. Hence
\[
    \det \mathcal{G}(\mathcal{I}^{(m)})\neq 0,
    \qquad m=1,\ldots,o.
\]

We now regard $b_t(1,o)$ as a variable and fix all remaining $\alpha(n-1)-1$ entries of $\bL_t$ arbitrarily in $\mathbb{F}_q$. Then $\det\mathcal{G}(\mathcal{I})$ is a polynomial in $b_t(1,o)$ of degree one, with nonzero linear coefficient
\[
    \pm \det\mathcal{G}(\mathcal{I}^{(1)}).
\]
Thus, this polynomial has at most one root in $\mathbb{F}_q$. Therefore, for each fixed assignment of the remaining $\alpha(n-1)-1$ entries, at most one value of $b_t(1,o)$ makes $\det\mathcal{G}(\mathcal{I})=0$. Hence, for a fixed independent set $\mathcal{I}$, at most
\[
    q^{\alpha(n-1)-1}
\]
matrices $\bL_t$ fail to preserve the independence of~$\mathcal{I}$.

There are at most
\[
    \binom{\alpha}{o}\binom{\alpha(n-1)}{B-o}
\]
index sets of the form $(B-o,o)$. Hence, the number of matrices that are non-$(B-o,o)$-preserving but are $(B-o',o')$-preserving for every $o'<o$ is at most
\[
    \binom{\alpha}{o}\binom{\alpha(n-1)}{B-o}q^{\alpha(n-1)-1}.
\]

Any matrix $\bL_t$ that fails to be $(B-o,o)$-preserving for some $o=1,2,\ldots,\alpha$ belongs to one of the above cases, corresponding to the smallest such value of $o$. Summing over $o$ therefore gives an upper bound on the number of bad matrices. Thus, a matrix that is $(B-o,o)$-preserving for every $o$ exists if
\[
    q^{\alpha(n-1)}
    >
    \left(
    \sum_{o=1}^\alpha
    \binom{\alpha}{o}\binom{\alpha(n-1)}{B-o}
    \right)
    q^{\alpha(n-1)-1}.
\]
%Equivalently, it suffices that
%\[
%    q
%    >
%    \sum_{o=1}^\alpha
%   \binom{\alpha}{o}\binom{\alpha(n-1)}{B-o}.
%\]

Using the standard convention that binomial coefficients outside their natural range are zero, we have
\[
    \sum_{o=0}^\alpha
    \binom{\alpha}{o}\binom{\alpha(n-1)}{B-o}
    =
    \binom{n\alpha}{B}.
\]
Therefore,
\[
    \sum_{o=1}^\alpha
    \binom{\alpha}{o}\binom{\alpha(n-1)}{B-o}
    =
    \binom{n\alpha}{B}-\binom{(n-1)\alpha}{B}.
\]
Thus, the stated field-size condition guarantees the existence of a matrix $\bL_t$ that preserves the independence of every independent set in $\mathscr{M}_t$.
\end{proof}

The dependence direction holds over any field and for every choice of local encoding coefficients; it requires no lower bound on the field size. The underlying fact is standard: a vertex cut in the signal flow graph confines the relevant global encoding vectors to a subspace of small dimension. This is the dependence side of the path representation of gammoids~\cite{Lindstrom73}, and the required separation is supplied by the vertex form of Menger's theorem~\cite[Ch.~13]{Welsh76}. We include a short self-contained proof.

\begin{theorem}
\label{thm: dependent}
Given the gammoid $\mathscr{M}_t=(\sV_t\cup\sV_{t+1},\mathscr{C}_t)$, for any choice of local encoding coefficients, if $\mathcal{V}(\mathcal{I})\subseteq \mathcal{V}_t\cup\mathcal{V}_{t+1}$ is dependent in $\mathscr{M}_t$, then $\mathcal{G}(\mathcal{I})$ is linearly dependent.
\end{theorem}

\begin{proof}
Since this proof does not rely on the specific coordinate labels of the vertices, we write $u$ for a vertex and $\be_u$ for its corresponding global encoding vector. The argument applies to every vertex, regardless of whether it is labeled as $v_t(i,j)$, $v_{t+1}(i',j')$, or otherwise.

Suppose that $\mathcal{V}(\mathcal{I})$ is dependent in $\mathscr{M}_t$. By definition, there is no linking of $\mathcal{V}(\mathcal{I})$ into $\mathcal{V}_{-1}$ in $G^{\mathrm{op}}$. Equivalently, there do not exist $|\mathcal{I}|$ mutually vertex-disjoint paths from $\mathcal{V}(\mathcal{I})$ to $\mathcal{V}_{-1}$ in $G^{\mathrm{op}}$. By the vertex form of Menger's theorem~\cite[Ch.~13]{Welsh76}, there exists a vertex set $\mathcal{S}$ with
\[
    |\mathcal{S}|<|\mathcal{I}|
\]
such that every path from any vertex of $\mathcal{V}(\mathcal{I})$ to $\mathcal{V}_{-1}$ in $G^{\mathrm{op}}$ passes through at least one vertex in $\mathcal{S}$.

Let
\[
    U := \operatorname{span}\{\be_w:\, w\in\mathcal{S}\}.
\]
Then
\[
    \dim U\leq |\mathcal{S}|<|\mathcal{I}|.
\]
We claim that, for every vertex $u$, if every directed path from $u$ to $\mathcal{V}_{-1}$ in $G^{\mathrm{op}}$ passes through $\mathcal{S}$, then $\be_u\in U$.

We prove the claim by induction on the stage of $u$. If $u\in\mathcal{S}$, then $\be_u\in U$ by definition. 

If $u$ is a source vertex at stage $-1$ and $u\notin\mathcal{S}$, then the trivial path from $u$ to itself avoids $\mathcal{S}$, contradicting the assumption. Hence, any source vertex satisfying the assumption must belong to $\mathcal{S}$, and the claim holds for stage $-1$.

Now consider a vertex $u$ at a later stage, and assume that the claim holds for all vertices in earlier stages. If $u\in\mathcal{S}$, then again $\be_u\in U$. Thus, suppose $u\notin\mathcal{S}$. Let $u'$ be an arbitrary in-neighbor of $u$ in the original signal flow graph $G$. Equivalently, there is an edge from $u$ to $u'$ in the opposite graph $G^{\mathrm{op}}$.

We show that every path from $u'$ to $\mathcal{V}_{-1}$ in $G^{\mathrm{op}}$ must pass through $\mathcal{S}$. Indeed, if there were a path from $u'$ to $\mathcal{V}_{-1}$ avoiding $\mathcal{S}$, then by prepending the edge from $u$ to $u'$ in $G^{\mathrm{op}}$, we would obtain a path from $u$ to $\mathcal{V}_{-1}$ avoiding $\mathcal{S}$, contradicting the assumption on $u$. Therefore, every path from $u'$ to $\mathcal{V}_{-1}$ passes through $\mathcal{S}$. Since $u'$ lies in an earlier stage, the induction hypothesis gives $\be_{u'}\in U$.

By the transfer relation in the signal flow graph, the global encoding vector at $u$ is a linear combination of the global encoding vectors associated with its in-neighbors in $G$:
\[
    \be_u=\sum_{u'} b(u',u)\be_{u'},
\]
where $b(u',u)\in\mathbb{F}_q$ is the coefficient of the edge from $u'$ to $u$ in $G$. Since every such $\be_{u'}$ lies in $U$ and $U$ is a subspace, it follows that $\be_u\in U$. This proves the claim.

Applying the claim to every vertex in $\mathcal{V}(\mathcal{I})$, we obtain that every vector in the indexed family $\mathcal{G}(\mathcal{I})$ belongs to $U$.
Since $|\mathcal{G}(\mathcal{I})|=|\mathcal{I}|$ while $\dim U\leq |\mathcal{S}|<|\mathcal{I}|$, $\mathcal{G}(\mathcal{I})$ is linearly dependent. This argument does not depend on the values of the local encoding coefficients, and therefore the conclusion holds for every choice of them.
\end{proof}

This completes the proof of Theorem~\ref{thm: linear_repre}. As stated in Section~\ref{sec: reduction}, the proof shows that the required finite-field size is independent of the stage index $t$. Moreover, at each stage, the local encoding coefficients can be chosen using only the current global encoding vectors $\mathcal{G}_t$ and the repair structure.

These two properties allow the theorem to be applied recursively from the initialization at stage $0$. Consequently, the linear representation over a fixed finite field persists under an unbounded sequence of repairs. We formalize this process in the proof of Corollary~\ref{coro: linear_repre}.

\begin{proof}[Proof of Corollary \ref{coro: linear_repre}]

We initialize $\mathcal{G}_0$ by choosing the global encoding vectors so that the vector matroid formed by $\mathcal{G}_0$ represents the initialization part of the signal flow graph. Equivalently, the initial global encoding vectors are chosen so that the required independence relations among the vertices at stage $0$ are realized as linear independence relations over $\FF{q}$. For instance, one may take the global encoding vectors to be the columns of a suitable $B\times n\alpha$ MDS generator matrix, viewed as vectors in $\FF{q}^B$, provided that the field size is large enough for such a matrix to exist. 

For the first repair, the coefficient-selection argument in the proof of Theorem~\ref{thm: linear_repre} applies directly to the initialized $\mathcal{G}_0$. Hence, the coefficients in $\bL_0$ can be chosen so that the vector matroid formed by $\mathcal{G}_0\cup\mathcal{G}_1$ is isomorphic to $\mathscr{M}_0$.

The field-size bound in Theorem~\ref{thm: linear_repre} is independent of the stage index. 
Indeed, under the parameter setting in~\eqref{eq: para}, the field-size bound in Theorem~\ref{thm: linear_repre} implies $q\geq n\alpha$, and hence an $[n\alpha,B]$ Reed-Solomon code may be used for the initialization. Therefore, over the same finite field, we may apply the theorem successively at stages $t=1,2,\ldots$. At each stage, Theorem~\ref{thm: linear_repre} provides local encoding coefficients such that the gammoid $\mathscr{M}_t$ is represented by the vector matroid formed by $\mathcal{G}_t\cup\mathcal{G}_{t+1}$.

Moreover, the construction of $\mathcal{G}_{t+1}$ depends only on $\mathcal{G}_t$ and the current graph structure. It does not depend on the global encoding vectors from earlier stages. Hence the representation can be extended indefinitely over the same fixed finite field. This proves the corollary.
\end{proof}

\section{Conclusion}
\label{sec:conclusion}
In this paper, we introduced a framework for constructing functional-repair regenerating codes using signal flow graphs and gammoids. The signal flow graph captures the evolution of the storage system under node failures and repairs, while the associated gammoids describe the combinatorial independence conditions required for data recovery. This viewpoint separates the code construction into two parts: first, the design of an unlabeled signal flow graph satisfying the required linking conditions; and second, the construction of a linear representation of the corresponding gammoids over a finite field.

Within this framework, we considered the functional repair of a single node failure in the regime $d=n-1$, where all surviving nodes participate as helpers. For each $\ell\in\{1,\ldots,k\}$, we constructed codes with the corresponding parameters on the storage-bandwidth tradeoff curve. Hence, the proposed construction attains every vertex point on the functional-repair tradeoff curve. Moreover, each repair is help-by-transfer and access-optimal: every helper reads exactly one stored packet and transfers it directly to the newcomer without performing local computation.

A key feature of the construction is its dynamic nature. Although the signal flow graph grows unboundedly as repairs proceed, the repair operation remains causal and finite-memory. In particular, the choice function at each stage depends only on a finite window of the failure history, rather than on the entire past sequence of failures. The local encoding coefficients are chosen over a fixed finite field whose size is independent of the number of repairs. Consequently, the construction supports an unbounded sequence of repairs, and the complexity of each repair does not grow with the number of preceding repairs.

Several directions remain open. One natural direction is to extend the signal-flow-graph and gammoid framework to functional-repair constructions for the case $d<n-1$, where only a subset of surviving nodes participates in repair. Another direction is to consider multiple-node failures and cooperative repair. The finite-field-size bound obtained in this paper is derived from an existence argument and is likely not tight; improving this bound or developing more explicit coefficient-selection algorithms would be of interest. Finally, adapting the framework to functional-repair regenerating codes under specific network structures, such as those considered in~\cite{mital2019practical,Mital}, is another promising direction for future work.

\appendices 

\section{Proof of Theorem \ref{thm: graph_property}}\label{apd: 3}

\begin{proof}[Proof of Theorem \ref{thm: graph_property}] 
By the definition of the gammoid $\mathscr{M}_t$, the set $\mathcal{U}$ is independent if and only if it can be linked into $\sV_{-1}$. Since the edges connect only consecutive stages, we introduce an auxiliary vertex subset $\mathcal{W}\subseteq \sV_t\setminus\mathcal{U}$ at stage $t$ that captures where the paths starting from $\mathcal{U}\cap\sV_{t+1}$ enter stage $t$. Recall that a linking is a set of vertex-disjoint paths. By the construction of the signal flow graph, the independence of $\mathcal{U}$ is therefore equivalent to the following claim.

\textbf{Claim.} There exists an auxiliary vertex subset $\mathcal{W}\subseteq \sV_{t}\setminus \mathcal{U}$ such that $\mathcal{U}\cap \sV_{t+1}$ can be linked onto $\mathcal{W}$, and $(\mathcal{U}\cap \sV_t)\cup \mathcal{W}$ can be linked into $\sV_{-1}$.

We first characterize the auxiliary sets $\mathcal{W}$ for which $\mathcal{U}\cap\sV_{t+1}$ can be linked onto $\mathcal{W}$. Since every path in such a linking starts at a vertex of $\mathcal{U}\cap\sV_{t+1}$ and reaches stage $t$ through a single incoming edge, the linking is determined by the edges terminating at the vertices of $\mathcal{U}\cap\sV_{t+1}$. We consider these vertices in two cases.

\begin{itemize}
    \item Assume $v_{t+1}(i,j)\in\mathcal{U}$ with $i\neq F_t$. By the construction of the edges, the unique carry-forward edge is the only edge terminating at $v_{t+1}(i,j)$, so the path through this vertex must enter stage $t$ at $v_t(i,j)$. Thus, $v_t(i,j)$ must belong to $\mathcal{W}$. Therefore, $$\mathcal{W}_1=\{v_t(i,j)\mid v_{t+1}(i,j)\in\mathcal{U}, i\neq F_t\}.$$
    \item Suppose $v_{t+1}(F_t,j)\in\mathcal{U}$. Only repair edges terminate at such a vertex, and each repair edge originates from a vertex of the form $v_t(i,p_t(i))$ with $i\neq F_t$. Since the paths are vertex-disjoint, distinct vertices $v_{t+1}(F_t,j)\in\mathcal{U}$ are matched to distinct vertices $v_t(i,p_t(i))\in\mathcal{W}$. Since the linking consists of vertex-disjoint paths, these vertices $v_t(i,p_t(i))$ must not belong to $\mathcal{W}_1$. The set of such vertices is $$\mathcal{W}_2\subseteq\{v_t(i,p_t(i))\mid i\neq F_t\}$$ of size $\bigl|\{v_{t+1}(F_t,j)\mid j\in[\alpha]\}\cap \mathcal{U}\bigr|$, satisfying $\mathcal{W}_1\cap\mathcal{W}_2=\emptyset$.
\end{itemize}

Let $\mathcal{W}=\mathcal{W}_1\cup\mathcal{W}_2$. The above discussion then shows that $\mathcal{U}\cap \sV_{t+1}$ can be linked onto $\mathcal{W}$.

Conversely, note that the choice of $\mathcal{W}_2$ does not depend on the specific nodes $i$. Therefore, any set $\mathcal{W}$ satisfying these two conditions yields a vertex-disjoint linking of $\mathcal{U}\cap\sV_{t+1}$ onto $\mathcal{W}$.

Therefore $\mathcal{U}$ is independent in $\mathscr{M}_t$ if and only if the two conditions hold simultaneously.
\end{proof}

\section{Proof of Theorem \ref{thm: alg}}
\label{apd: 1}
\begin{proof}
   We prove by induction on $t$ that the signal flow graph generated by Algorithm~\ref{alg1} is admissible. Suppose the choice functions at stages $0, 1, \ldots, t-1$ are admissible, i.e., for any node $i$ and any stages $a < b \leq t-1$ with $|\sF{a,b}| \leq \alpha$ and $i \notin \sF{a,b}$, $p_a(i) = p_b(i)$ if and only if $F_a = F_b$. It suffices to show that for every stage $a < t$ and every node $i$ satisfying $|\sF{a,t}| \leq \alpha$ and $i \notin \sF{a,t}$, the value $p_t(i)$ given by the algorithm satisfies $p_t(i) = p_a(i)$ if and only if $F_t = F_a$.

    Let $i$ be an arbitrary node with $i \neq F_t$. We consider four cases according to the choice of the stage $c$.
\begin{itemize}
    \item Suppose $c = -1$. It then follows that $F_t \notin \sF{0,t-1}$ and $|\sF{0,t-1}| < \alpha$. 
    
    The algorithm assigns $p_t(i) \in [\alpha] \setminus \sP{0,t-1}(i)$, thus $p_t(i) \neq p_a(i)$ for every stage $a < t$. Since $F_t \notin \sF{0,t-1}$, we have $F_t \neq F_a$ for every such $a$. Thus the admissibility condition holds.

    \item Suppose $F_c = F_t$. It then follows that $F_t \notin \sF{c+1,t-1}$, $i \notin \sF{c,t}$, $|\sF{c,t-1}| \leq \alpha$, and the algorithm assigns $p_t(i) = p_c(i)$. 
    
    Consider any stage $a < t$ with $|\sF{a,t}| \leq \alpha$. If $a \leq c$, then $\sF{a,c} \subseteq \sF{a,t}$ and thus $|\sF{a,c}| \leq \alpha$. If $c < a < t$, then $\sF{c,a} \subseteq \sF{c,t}$, thus $|\sF{c,a}| \leq \alpha$. In either case, the induction hypothesis gives $p_a(i) = p_c(i)$ if and only if $F_a = F_c$. Since $F_c = F_t$ and $p_c(i) = p_t(i)$, it follows that $p_a(i) = p_t(i)$ if and only if $F_a = F_t$, as required.

\item Suppose $F_c=i$. It then follows that $F_t\notin\sF{c,t-1}$ and $|\sF{c,t-1}|\leq \alpha$. 

Since $i\in\sF{a,t}$ for stage $a<c$, it suffices to consider stage $a$ with $c<a<t$. The algorithm gives $p_t(i)\in[\alpha]\setminus\sP{c+1,t-1}(i)$, thus $p_t(i)\neq p_a(i)$ for every stage $a$. Since $F_t\neq F_a$ also holds for every $c<a<t$, the admissibility follows.

\item In the remaining case, $F_c\notin\sF{c+1,t}$, $F_t\notin\sF{c,t-1}$ and $|\sF{c,t-1}|=|\sF{c+1,t}|=\alpha$. 

Since $F_c\notin\sF{c+1,t-1}$ and $|\sF{c,t-1}|=\alpha$, we have $|\sF{c+1,t-1}|=\alpha-1$, so the induction hypothesis gives $|\sP{c+1,t-1}(i)|=\alpha-1$ and $p_c(i)\notin\sP{c+1,t-1}(i)$, and thus $p_c(i)$ is the unique element of $[\alpha]\setminus\sP{c+1,t-1}(i)$.
The algorithm sets $p_t(i)=p_c(i)$, thus $p_t(i)\notin\sP{c+1,t-1}(i)$. For any stage $a$ with $|\sF{a,t}|\leq\alpha$, the condition $|\sF{c+1,t}|=\alpha$ implies $a>c$, therefore $p_t(i)\neq p_a(i)$ and $F_t\neq F_a$, i.e., the admissibility holds. 
    \end{itemize}
\end{proof}

\bibliographystyle{IEEEtran}
\bibliography{DStorage}

\end{document}